\documentclass[useAMS,usenatbib]{mnras}

\usepackage{newtxtext,newtxmath}

\usepackage[T1]{fontenc}
\usepackage{ae,aecompl}

\usepackage{graphicx}	
\usepackage{amsmath}	
\usepackage{amssymb}	
\usepackage{epstopdf}
\usepackage{subcaption}
\usepackage[usenames, dvipsnames]{color}

\usepackage{etoolbox}
\makeatletter
\patchcmd\@combinedblfloats{\box\@outputbox}{\unvbox\@outputbox}{}{%
   \errmessage{\noexpand\@combinedblfloats could not be patched}%
}%
 \makeatother

\title[Pure-helium 3D model atmospheres of white dwarfs]{Pure-helium 3D model atmospheres of white dwarfs}

\author[Cukanovaite et al.]{E. Cukanovaite$^1$\,{\Huge \footnotemark},
P.-E. Tremblay$^1$, B. Freytag$^2$, H.-G. Ludwig$^3$,
  \newauthor{and P. Bergeron$^4$}
  \\
$^{1}$ Department of Physics, University of Warwick, Coventry CV4 7AL, UK \\ 
$^{2}$ Department of Physics and Astronomy, Uppsala University, Box 516, 751 20 Uppsala, Sweden \\
$^{3}$ Zentrum f\"ur Astronomie der Universit\"at Heidelberg, Landessternwarte, K\"onigstuhl 12, 69117 Heidelberg, Germany \\
$^{4}$ D\'epartement de Physique, Universit\'e de Montr\'eal, C.P. 6128, Succ. Centre-Ville, Montr\'eal, QC H3C 3J7, Canada 
}

\date{Accepted XXX. Received YYY; in original form ZZZ}

\pubyear{2018}

\begin{document}
\label{firstpage}
\pagerange{\pageref{firstpage}--\pageref{lastpage}}
\maketitle

\begin{abstract}

We present the first grid of 3D simulations for the pure-helium atmospheres of DB white dwarfs. The simulations were computed with the CO$^5$BOLD radiation-hydrodynamics code and cover effective temperatures and surface gravities between 12\,000~K~$\lesssim$~$T_{\rm{eff}}$~$\lesssim$~34\,000~K and 7.5~$\leq$~$\log{g}$~(cgs units)~$\leq$~9.0, respectively. In this introductory work, synthetic spectra calculated from the 3D simulations are compared to appropriate 1D model spectra under a differential approach. This results in the derivation of 3D corrections for the spectroscopically derived atmospheric parameters of DB stars with respect to the 1D ML2/$\alpha$ = 1.25 mixing-length parameterisation. No significant $T_{\rm{eff}}$ corrections are found for the V777 Her instability strip region, and therefore no 3D revision is expected for the empirical blue and red edges of the strip. However, large $\log{g}$ corrections are found in the range 12\,000\,K < $T_{\rm{eff}}$ < 23\,000\,K for all $\log{g}$ values covered by the 3D grid. These corrections indicate that 1D model atmospheres overpredict $\log{g}$, reminiscent of the results found from 3D simulations of pure-hydrogen white dwarfs. The next step will be to compute 3D simulations with mixed helium and hydrogen atmospheres to comprehend the full implications for the stellar parameters of DB and DBA white dwarfs.

\end{abstract}

\begin{keywords}
white dwarfs --  stars: atmospheres -- convection -- hydrodynamics -- techniques: spectroscopic
\end{keywords}

\footnotetext{E-mail: E.Cukanovaite@warwick.ac.uk}


\section{Introduction}

Most white dwarfs have hydrogen dominated atmospheres as a result of gravitational settling \citep{schatzman48} due to their large surface gravities. The atmospheres of a significant number of degenerate stars is however dominated by helium, which is understood to be the consequence of late thermal pulses experienced by post-asymptotic giant branch (AGB) progenitors \citep{althaus05, werner06}. These white dwarfs are classified with spectral type DO if He~II lines dominate, DB if He~I lines dominate, or DC if the $T_{\rm eff}$ is too low to allow any optical transition. Subtypes such as DBA or DBZ are also employed to designate additional hydrogen or metal lines, respectively. About 20\% of white dwarfs in magnitude limited samples \citep[e.g. SDSS;][]{kleinman13,kepler15} are of DB or DO spectral types. For volume-complete samples where the white dwarf luminosity function peaks at much cooler temperatures, the fraction of helium-dominated atmospheres is as large as 50\% \citep{gia12}. This increase of helium-rich stars below $T_{\rm eff} \sim$ 10\,000~K is likely due to convective mixing events in hydrogen-line (DA) white dwarfs, resulting in their thin hydrogen blanket being fully mixed-in with the underlying helium layer \citep{tremblay08}.

For the majority of DB and DBA white dwarfs, the spectroscopic technique, which compares the observed line profiles to predictions from model spectra \citep[][henceforth BW11]{bergeron_db_2011}, is used to determine their atmospheric parameters ($T_{\rm eff}$, $\log g$, and H/He). These parameters coupled with evolutionary models allow for the determination of white dwarf masses and ages. While DB white dwarfs are not quite as frequent as DA or DC spectral types, their parameters are still essential to understand the local stellar formation history \citep{kalirai_2012_ages_clusters,tremblay14}, the late thermal pulses in post-AGB progenitors \citep{reindl14b,reindl14a}, and the fraction of primordial hydrogen in white dwarfs \citep[BW11;][]{koester_kepler_2015,rolland18}. Furthermore, a large fraction of white dwarfs polluted by asteroids and planetary debris \citep{veras16} have helium-dominated atmospheres \citep{kleinman13}. This is expected from the much larger diffusion timescales for the denser helium atmospheres \citep{paquette86a,paquette86b,koester09,fontaine15}. As a consequence, DB white dwarfs are important objects for the understanding of post-main-sequence planetary system evolution, and in particular the detection of water-rich asteroids \citep{farihi13,raddi15,gentile17}. 

In the last two decades, detailed spectroscopic analyses of DB and DBA white dwarfs have resulted in exquisite mass-$T_{\rm eff}$ distributions, and sophisticated 1D model atmosphere and spectral synthesis codes which incorporate detailed line broadening schemes (see, e.g., \citealt{beauchamp_stark_profiles_1997}; \citealt{voss_2007_db_dba_survey}; BW11; \citealt{koester_kepler_2015}). The importance of hydrogen in the helium-dominated atmosphere white dwarfs has also been made clear. These studies have determined that the fraction of DBAs in DB/DBA samples is up to 75\%, and that perhaps all DB white dwarfs have traces of hydrogen with the abundance too low to cause observable spectral features \citep{koester_kepler_2015}, illustrating the close link between the two spectral classifications. The presence of hydrogen, even if not observed, can significantly affect the derived $T_{\rm{eff}}$, especially for the boundaries of the V777 Her instability strip \citep{beauchamp_1999_v777_dba}, the region where pulsating DB/DBA white dwarfs are found.

Some issues remain in the spectroscopic analyses of DB white dwarfs. One such problem is observed at $T_{\rm{eff}} < 16\,000$~K, where the spectroscopically derived surface gravities are significantly higher than the predictions of evolutionary models, possibly due to incomplete treatment of line broadening by neutral helium \citep [BW11;][]{koester_kepler_2015}. \cite{beauchamp_1996_van_der_waals} have shown that better treatment of the van der Waals broadening implemented from \cite{deridder_van_der_waals_broad_low_t_1976} does lower the surface gravities at low effective temperatures, yet the authors find that gravity is very sensitive to the exact treatment of this broadening. A similar high-$\log{g}$ problem was known for DA white dwarfs \citep{bergeron_1990_highloggda} for $T_{\rm{eff}} <$\,12\,000~K, a temperature which corresponds to the onset of convective energy transfer in the photosphere of DAs. This problem was solved by computing the first-ever 3D model atmospheres of DA white dwarfs \citep{tremblay_2013_3dmodels,tremblay_2013_spectra} and corresponding 3D synthetic spectra, confirming the long-standing suspicion that convection is modelled in a too-crude manner in 1D model atmospheres.
 
Helium-atmosphere white dwarfs develop superficial convection zones at temperatures as large as 50\,000~K and thus all currently known DB stars must rely on convective model atmospheres. Consequently, the high-$\log{g}$ problem for cool DB white dwarfs cannot be related to the onset of convection, but it could be caused by changes in the properties of convection that 1D models do not consider. Therefore, it is of great interest to look at the predictions of 3D DB model atmospheres, especially because of the success of modelling DA white dwarfs. In addition to the derivation of masses and ages that are likely to be more precise, the sizes of the convection zones and overshoot regions \citep{tremblay_2015_mlt_cal,kupka18} are of particular importance, since they determine the total mass of hydrogen or accreted metals in DBA and DBZ white dwarfs, respectively. Another critical aspect is the revision of the spectroscopic parameters to determine the empirical edges of the V777 Her instability strip and the connection to asteroseismic models \citep{vangrootel17}.

In 1D white dwarf model atmospheres, convection is treated using the ML2 version (\citealt{tassoul_1990_ml2}; see also Table 1 of \citealt{tremblay_2013_3dmodels}) of the mixing-length approximation \citep{bohm_vitense_1958_mlt}. The mixing-length parameter, $\alpha$, is usually set to 1.25 for DB stars \citep[BW11;][]{koester_kepler_2015}. BW11 derived this value by looking at possibly unphysical clumping in the $\log{g}$-$T_{\rm{eff}}$ distribution arising from the different $\alpha$ values and from the calibration of the $T_{\rm{eff}}$ derived from fits of optical and UV spectra. Although ML2/$\alpha = 1.25$ performed reasonably well in both tests, BW11 suggested that an improvement needs to be made in the treatment of convective transport itself, which is exactly what 3D models can provide. Contrary to the mixing length approximation which requires the variation of $\alpha$ to reproduce even the basic properties of astrophysical observations, 3D numerical simulations do not require such fine tuning.

In this paper, we present the first 3D radiation-hydrodynamics simulations for pure-helium white dwarf atmospheres. From these simulations we compute synthetic DB spectra that we compare with 1D spectra and selected published observations. We postpone the full spectroscopic analysis of existing data sets until grids of 3D model atmospheres with mixed He/H compositions become available. Nevertheless, our predictions with a pure-helium equation-of-state (EOS) will be useful to interpret the independent stellar parameters ($T_{\rm eff}$, radius, luminosity) revealed from the {\it Gaia} Data Release 2 \citep{tremblay17,bedard17,hollands18}. Our study is restricted to the atmospheric properties of DB white dwarfs; the calibration of the mixing-length theory for structure calculations will be presented elsewhere. In Section~\ref{sec:3d_sec}, the numerical setup of the 3D simulations is explained and some brief description of the structural differences between the 3D and 1D convection models is given. The calculation of synthetic spectra for both 3D and 1D structures is explained in Section~\ref{sec:model_spectra}. Our proposed 3D corrections are presented and discussed in Section~\ref{sect:3d_corr}. We conclude in Section~\ref{sec:conc}.

\section{Model atmospheres} 
\label{sec:3d_sec}

\subsection{Numerical setup for CO$^5$BOLD simulations} \label{sec:numsetup}

\begin{figure}
	\includegraphics[width=\columnwidth]{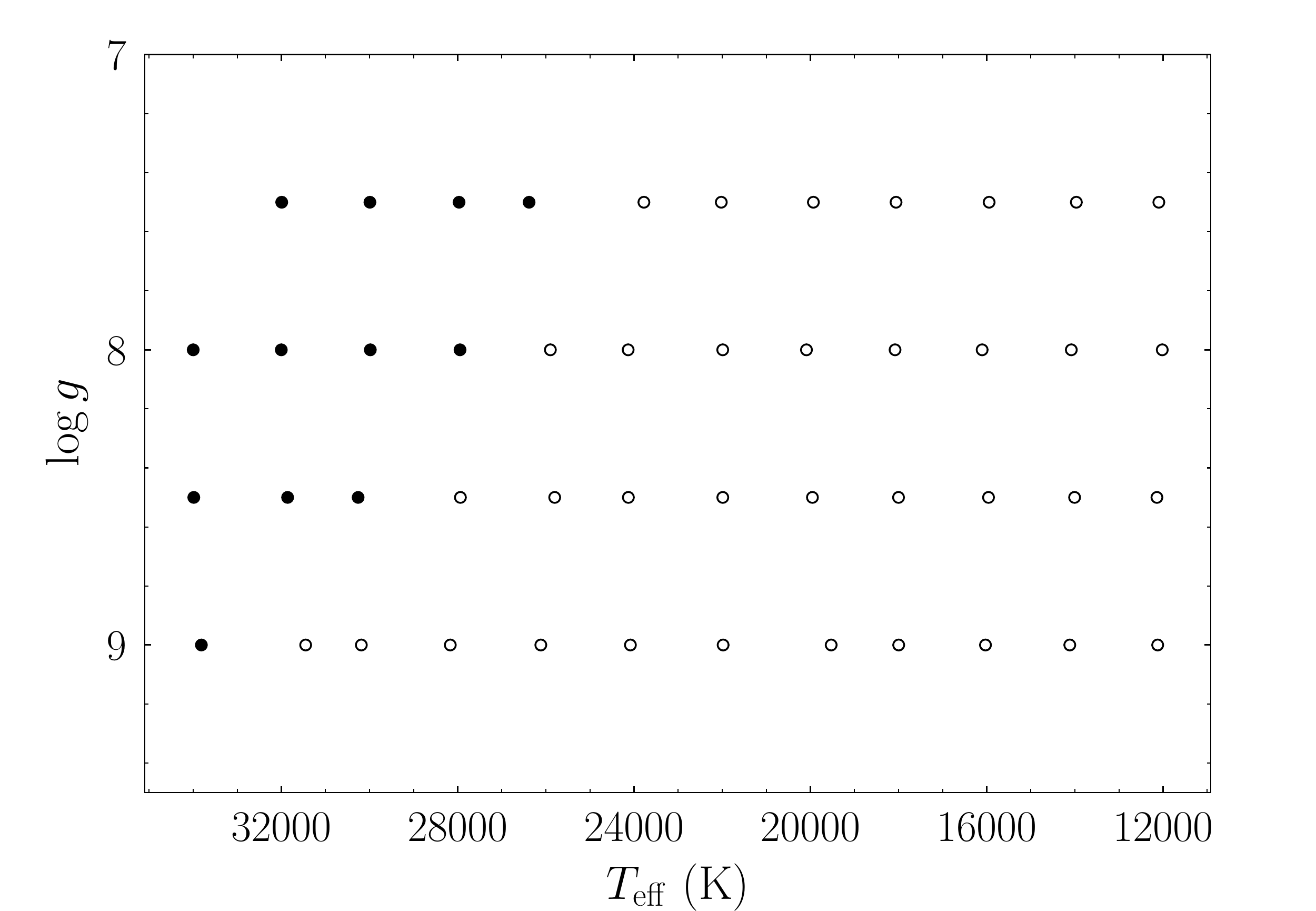}
    \caption{The $T_{\rm eff}$ and $\log g$ of our 3D model atmospheres with pure-helium compositions. Open and filled circles denote models with bottom boundaries that are open and closed to convective flows, respectively.}
    \label{fig:models}
\end{figure}

The CO$^5$BOLD radiation-hydrodynamics code \citep{freytag2012_cobold} was used to compute 47 3D DB model atmospheres with $\log{g}$ ranging between 7.5 and 9.0 in steps of 0.5 dex, and $T_{\rm{eff}}$ between 12\,000~K and 34\,000~K in steps of around 2000 K. Our grid is illustrated in Fig.~\ref{fig:models} and presented in Table~\ref{tab:3d_models}. When computing 3D simulations $\log{g}$ is an input parameter. However, the effective temperature is not, and is instead recovered after the model is calculated, from the spatially and temporally averaged emergent stellar flux. The temporal average is restricted to the last one-fourth of the simulations. This results in the unevenly spaced $T_{\rm{eff}}$ values of our 3D models.

\begin{table}
	\centering
	\caption{Selected parameters of the pure-helium 3D model atmospheres. $T_{\rm{eff}}$ is calculated from the spatially and temporally averaged emergent stellar flux and $\delta I_{\rm rms}/\langle I \rangle$ is the relative bolometric intensity contrast.}
	\label{tab:3d_models}
	\begin{tabular}{lcccccr} 
		\hline
		$\log{g}$ & $T_{\rm{eff}}$ & Box size & Time & $\delta I_{\rm rms}/\langle I \rangle$ \\
		 & (K) & (km $\times$ km $\times$ km) & (stellar s) & (\%) \\
		\hline
		7.5 & 12098 & 1.22$\times$1.22$\times$0.58 & 31.6 & 3.6 \\
        7.5 & 13969 & 1.98$\times$1.98$\times$0.67 & 31.6 & 8.9 \\
        7.5 & 15947 & 2.86$\times$2.86$\times$1.19 & 31.6 & 16.4 \\
        7.5 & 18059 & 6.09$\times$6.09$\times$1.46 & 31.6 & 21.3 \\
        7.5 & 19934 & 11.96$\times$11.96$\times$2.39 & 31.6 & 23.4 \\
        7.5 & 22023 & 21.75$\times$21.75$\times$4.51 & 31.6 & 25.5 \\
        7.5 & 23778 & 23.96$\times$23.96$\times$4.78 & 31.6 & 24.3 \\
        7.5 & 26382 & 37.47$\times$37.47$\times$10.88 & 31.6 & 22.9 \\
        7.5 & 27970 & 31.22$\times$31.22$\times$10.77 & 15.0 & 17.5 \\
        7.5 & 29992 & 31.22$\times$31.22$\times$11.86 & 20.0 & 9.4 \\
        7.5 & 31993 & 33.48$\times$33.48$\times$14.00 & 8.0 & 4.9 \\  
        \hline
        8.0 & 12020 & 0.70$\times$0.70$\times$0.10 & 10.0 & 2.1 \\
        8.0 & 14083 & 0.79$\times$0.79$\times$0.24 & 10.0 & 6.0 \\
        8.0 & 16106 & 0.94$\times$0.94$\times$0.18 & 10.0 & 11.9 \\
        8.0 & 18081 & 1.23$\times$1.23$\times$0.35 & 10.0 & 17.0 \\
        8.0 & 20090 & 2.00$\times$2.00$\times$0.58 & 10.0 & 19.4 \\
        8.0 & 21989 & 5.19$\times$5.19$\times$0.97 & 10.0 & 22.3 \\
        8.0 & 24135 & 8.62$\times$8.62$\times$1.41 & 10.0 & 23.8 \\
        8.0 & 25899 & 8.62$\times$8.62$\times$1.56 & 10.0 & 21.1 \\
        8.0 & 27948 & 17.69$\times$17.69$\times$3.04 & 10.0 & 20.6 \\
        8.0 & 29983 & 12.63$\times$12.63$\times$3.50 & 10.0 & 19.7 \\
        8.0 & 32002 & 12.63$\times$12.63$\times$3.28 & 10.0 & 14.8 \\
        8.0 & 33999 & 12.63$\times$12.63$\times$3.42 & 10.0 & 7.9 \\
		\hline
        8.5 & 12141 & 0.25$\times$0.25$\times$0.05 & 3.2 & 1.5 \\
        8.5 & 14009 & 0.25$\times$0.25$\times$0.04 & 3.2 & 3.6 \\
        8.5 & 15961 & 0.34$\times$0.34$\times$0.05 & 3.2 & 7.6 \\
        8.5 & 18002 & 0.39$\times$0.39$\times$0.13 & 3.2 & 12.6 \\
        8.5 & 19955 & 0.60$\times$0.60$\times$0.20 & 3.2 & 15.5 \\
        8.5 & 21988 & 1.03$\times$1.03$\times$0.26 & 3.2 & 17.8 \\
        8.5 & 24130 & 1.78$\times$1.78$\times$0.37 & 3.2 & 22.1 \\
        8.5 & 25801 & 2.37$\times$2.37$\times$0.44 & 3.2 & 22.3 \\
        8.5 & 27939 & 2.53$\times$2.53$\times$0.59 & 3.2 & 20.6 \\
        8.5 & 30259 & 4.53$\times$4.53$\times$1.23 & 3.2 & 20.4 \\
        8.5 & 31859 & 4.53$\times$4.53$\times$1.23 & 3.2 & 19.7 \\
        8.5 & 33987 & 4.53$\times$4.53$\times$0.98 & 3.2 & 17.6 \\
        \hline
        9.0 & 12124 & 0.06$\times$0.06$\times$0.01 & 1.0 & 0.8 \\
        9.0 & 14118 & 0.07$\times$0.07$\times$0.01 & 1.0 & 2.3 \\
        9.0 & 16030 & 0.11$\times$0.11$\times$0.02 & 1.0 & 5.0 \\
        9.0 & 17999 & 0.12$\times$0.12$\times$0.03 & 1.0 & 8.7 \\
        9.0 & 19530 & 0.12$\times$0.12$\times$0.03 & 1.0 & 11.2 \\
        9.0 & 21981 & 0.20$\times$0.20$\times$0.07 & 1.0 & 13.6 \\
        9.0 & 24084 & 0.39$\times$0.39$\times$0.10 & 1.0 & 17.2 \\
        9.0 & 26116 & 0.76$\times$0.76$\times$0.13 & 1.0 & 20.6 \\
        9.0 & 28169 & 0.76$\times$0.76$\times$0.16 & 1.0 & 20.6 \\
        9.0 & 30187 & 0.86$\times$0.86$\times$0.20 & 1.0 & 17.4 \\
        9.0 & 31449 & 0.86$\times$0.86$\times$0.20 & 1.0 & 17.2 \\
        9.0 & 33815 & 1.43$\times$1.43$\times$0.39 & 1.0 & 19.1 \\
        \hline
	\end{tabular}
\end{table}
 
Apart from the EOS and opacities, our computational setup is the same as that used for DA white dwarfs \citep{tremblay_2013_3dmodels,tremblay_2013_spectra}. In brief, each model is computed in a box of $150 \times 150 \times 150$ Cartesian ($x \times y \times z$) grid points, where $z$-axis represents the vertical extent of the simulation, pointing from the interior to the exterior of the white dwarf. We use the local box-in-a-star CO$^5$BOLD setup (see \citealt{freytag2012_cobold} for more detail), where only a section of the atmosphere is modelled. Each simulation has periodic side boundaries and a top boundary that is open to material and radiative flows. Radiative transfer is solved everywhere in the simulation with the diffusion approximation applied at the bottom boundary. Depending on the size of the convection zone, the bottom boundary is either open or closed to convective flows. In Fig.~\ref{fig:models}, the models with open and closed bottom boundaries are indicated by open and filled circles, respectively. For most of the effective temperatures considered in this study, the convection zones are too large to fit inside the vertical extent of the simulation. Instead, enough of the convection zone is included vertically such that convection becomes adiabatic near the bottom, ensuring that the inflowing material at the open bottom boundary can be described by adiabatic convection. At the highest effective temperatures, the convection zones become small enough to fit within the simulations and the bottom boundaries are closed to convective flows but left open to radiation. In those cases, the vertical velocities are forced to go to zero and the radiative flux is prescribed to the desired total energy flux. The bottom layer for all models is around $\log{\langle\tau_{\rm{R}}\rangle} = 3 $, where $\langle \tau_{\rm{R}} \rangle$ is Rosseland optical depth averaged over space and time. Some closed boundary models were extended deeper to include a larger overshoot region. The top $\langle\tau_{\rm{R}}\rangle$ value varies from model to model, ranging from $-8.5 \lesssim \log{\langle\tau_{\rm{R}}\rangle} \lesssim -4.8 $, with all simulations covering the line forming region. Horizontally, along the $x$-axis, at least 4 convective cells are included to both resolve surface granulation and obtain meaningful mean properties. 

All simulations cover more than 4.5 pressure scale heights vertically, with the majority being more than 10 pressure scale heights deep. It has, however, been shown that the vertical boundaries can impact atmospheric layers within two pressure scale heights \citep{Grimm_Strele_2pressurescale}. This might affect the spectral properties for our shallowest models and we dedicate Section~\ref{sec:numerics} to this discussion.

The simulations cover a stellar time of a minimum of 60 turnover timescales at $\langle \tau_{\rm R} \rangle = 1$. We have confirmed that our models are relaxed in the last quarter of the simulation by monitoring total flux as a function of depth over time (including outgoing flux at the top). In all cases systematic variations within that time frame were less than the statistical noise due to periodic waves and the finite number of convective cells in our simulations. Convergence of the velocity field was also reached for all cases but the lowest effective temperature models, where the velocity field is still not in equilibrium in the uppermost layers ($\langle\tau_{\rm{R}}\rangle < -3$). As stated in \cite{tremblay_2013_3dmodels,tremblay_2013_spectra}, the upper layers never reach radiative equilibrium owing to very large Peclet numbers, and instead the entropy gradient slowly converges to a near-adiabatic structure due to the weak convective overshoot.

In the following we employ $\langle \rm{3D} \rangle$ averages which are derived from spatial and temporal averages of the 3D simulations over constant $\tau_{\rm{R}}$ surfaces. 12 snapshots in the last one-fourth of each simulation were used. These average structures are useful both for a simple comparison with 1D structures and as inputs for $\langle \rm{3D} \rangle$ spectral synthesis in Section~\ref{sec:model_spectra}. In Section~\ref{sec32} we report on the possibility of performing full 3D spectral synthesis instead of using $\langle \rm{3D} \rangle$ structures.

\subsection{Input microphysics and 1D LHD code}\label{sec:lhd}

\begin{figure*}
	\includegraphics[width=2\columnwidth]{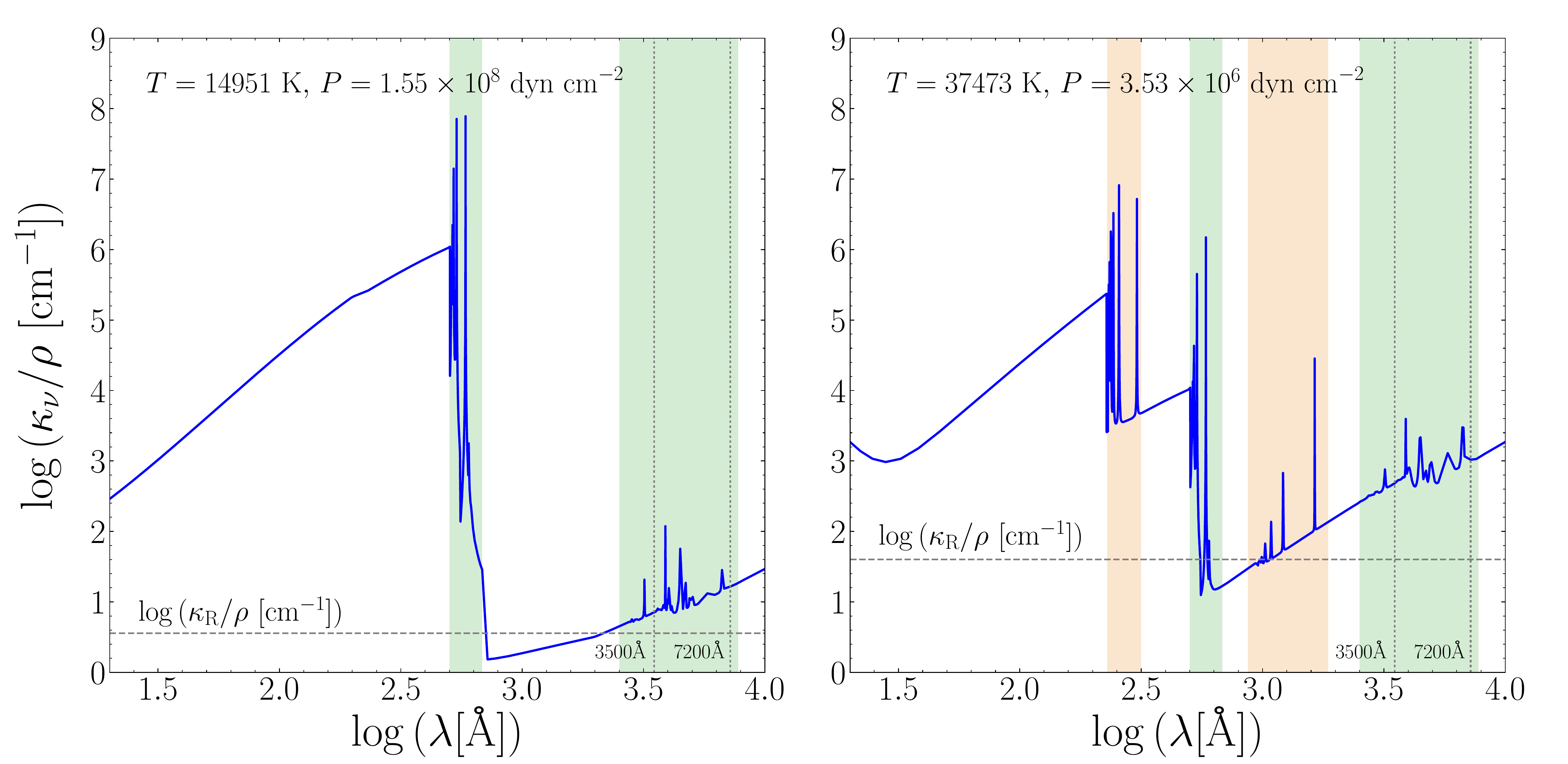}
    \caption{The total opacity as a function of the logarithm of wavelength for representative layers of two BW11 $\log{g} = 8.0$ models at $T_{\rm{eff}} = 14\,000$~K (left) and $T_{\rm{eff}} = 34\,000$~K (right). We have selected reference temperature and pressure values (as indicated on the panels) that correspond to the plasma conditions at $\tau_{\rm{R}} = 1$. He~I and He~II line opacities are indicated by green and red colour regions. The Rosseland mean opacity (dashed grey) and the line region used for the derivation of the 3D corrections (dotted grey) are also shown.}
    \label{fig:opacities}
\end{figure*}

\begin{figure*}
	\includegraphics[width=2\columnwidth]{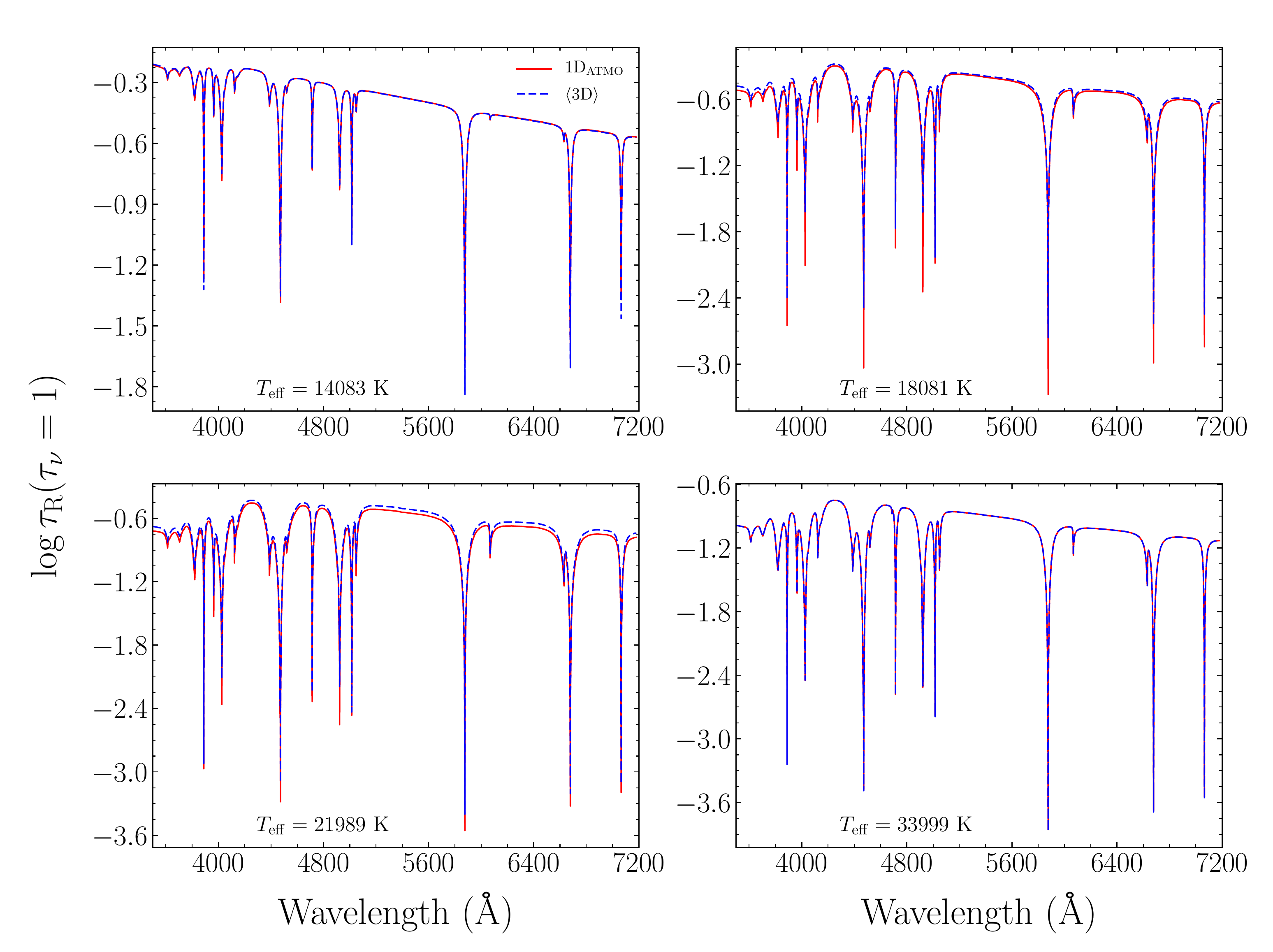}
    \caption{Atmospheric line forming regions for 1D ATMO (solid red) and $\langle \rm{3D} \rangle$ (dashed blue) spectra as defined by $\tau_{\rm{R}}( \tau_{\nu} = 1 )$, where the plasma becomes optically thin for photons of frequency $\nu$.}
    \label{fig:tau_ross_comp}
\end{figure*}

Microphysics in the form of EOS and opacity tables, are input parameters for CO$^5$BOLD. The EOS and opacity tables have been pre-calculated from reference 1D models, which in our case are the standard 1D DB model atmospheres of BW11 calculated using their 1D atmosphere code, hereafter referred to as ATMO. For reference, Fig.~\ref{fig:opacities} shows the opacity as a function of wavelength for the photosphere of two BW11 $\log{g} = 8.0$ models. For helium-dominated compositions, ATMO includes improved Stark profiles of neutral helium calculated by~\cite{beauchamp_stark_profiles_1997} and the free-free absorption coefficient of negative helium ions of~\cite{john_neg_ab_coeff_neg_he_ion_1994}. This 1D code uses 1745 frequencies to solve radiative transfer for the atmospheric structure, which is a procedure that is computationally impossible to replicate for 3D models and so opacity binning is used instead. This method relies on sorting the frequency-dependent opacities based on their corresponding frequency-dependent optical depth, $\tau_{\nu}$, such that $\tau_{\nu}$ (and therefore the associated opacity), will fall in an $i$-th bin defined by two values of $\tau_{\rm{R}}$, i.e. [$\tau_{\rm{R}}^{i-1}$, $\tau_{\rm{R}}^{i}$], if 
\begin{equation}
\tau_{\rm{R}}^{i-1} \geqslant \tau_{\rm{R}}( \tau_{\nu} = 1 ) > \tau_{\rm{R}}^{i},
\end{equation}
(\citealt{nordlund_1982_opac_binning}; \citealt{ludwig_1994_op_binning}; \citealt{vogler_2004_op_binning}). The optical depth at which plasma becomes optically thin for photons of frequency $\nu$ is defined by $\tau_{\rm{R}}( \tau_{\nu} = 1 )$ and this is shown in Fig.~\ref{fig:tau_ross_comp} for selected 1D and mean 3D, hereafter $\langle \rm{3D} \rangle$, spectra (see Section~3) at $\log{g} = 8.0$. Each opacity table has been computed with 10 band-averaged opacity bins with boundaries at $\log{\tau_{\rm{R}}}$ = [99.0, 0.25, 0.0, $-$0.25, $-$0.5, $-$1.0, $-$1.5, $-$2.0, $-$3.0, $-$4.0, $-$5.0]. We note that due to interpolation issues we did not include the extremely strong far-UV opacities whenever they were assigned to the missing $\log{\tau_{\rm{R}}}=$ [$-$5.0,$-$99.0] opacity bin. As Fig.~\ref{fig:opacities} shows, at low effective temperatures He~I bound-free and He~I lines from the ground level provide the far-UV opacities. At high effective temperatures He~II bound-free and He~II line opacities also contribute. These frequencies are fully opaque to light everywhere in the simulations and very little flux is transported at such short wavelengths in the photosphere, therefore this missing opacity has little impact on the resulting temperature and pressure stratifications that are used for spectral synthesis.

Another important difference between standard 1D structures and our 3D simulations comes from CO$^5$BOLD treating scattering as true absorption, again, due to current numerical limitations. One therefore may argue that any 3D effects we observe when comparing our 3D models with 1D ATMO structures are due to approximations with opacity tables, scattering and even the missing opacities mentioned earlier. To test this hypothesis, the ATMO structures were compared with stratifications calculated using a different 1D code called LHD~\citep{caffau_2007_lhd}. The LHD code treats microphysics by employing the same input tables as those used in CO$^5$BOLD, considers scattering as true absorption, and has been modified to rely on a mixing length parameterisation of ML2/$\alpha = 1.25$. \citet{tremblay_2013_3dmodels} have shown, from the comparison of pure-hydrogen structures, that differences between the LHD and ATMO codes are small apart from the input microphysics. Consequently, any difference observed between them in the case of pure-helium composition would likely be caused by approximations in the microphysics.

\begin{figure*}
	\includegraphics[width=2\columnwidth]{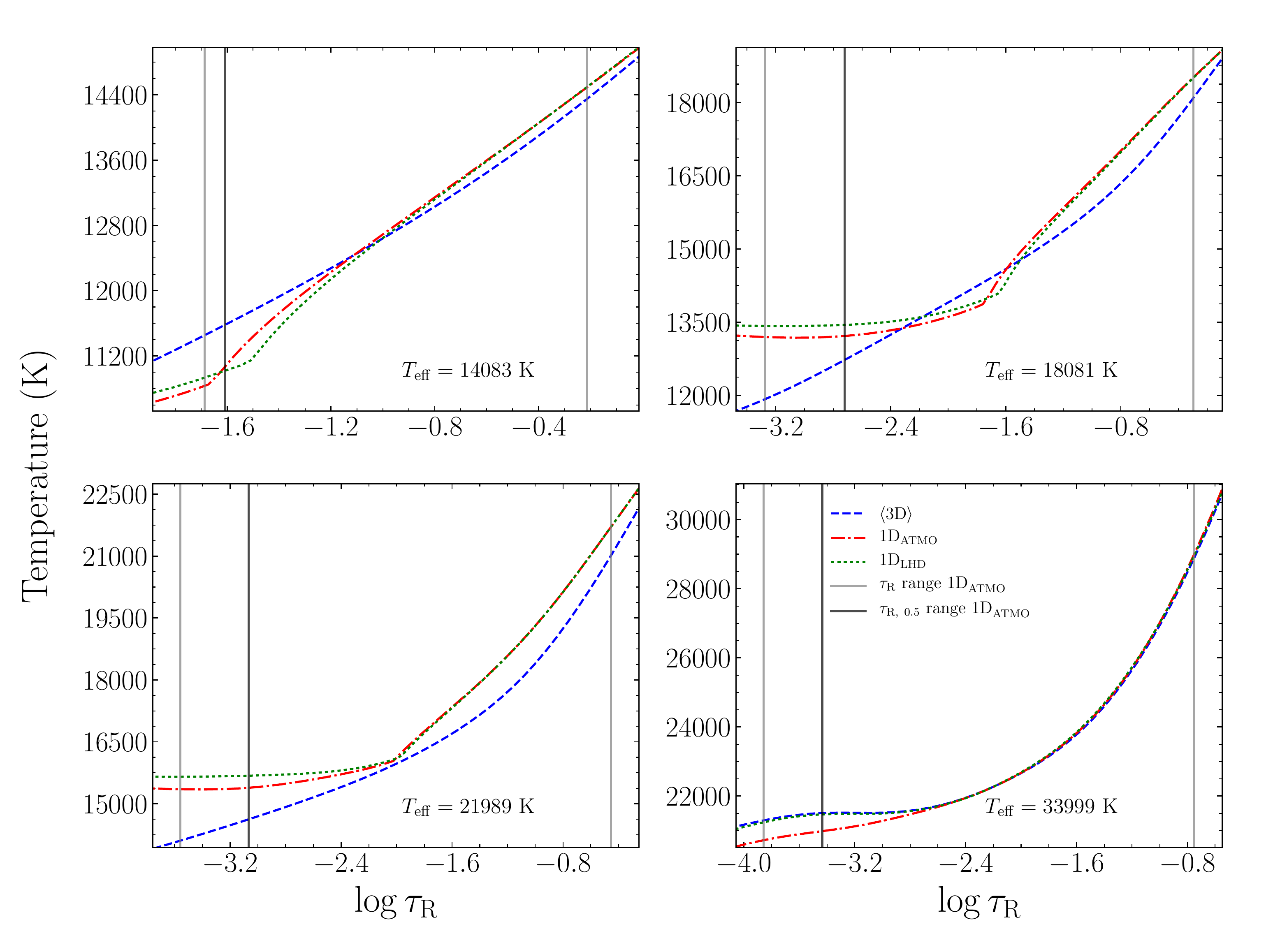}
    \caption{Temperature stratifications of spectral line forming regions for $\langle \rm{3D} \rangle$ (dashed blue), 1D ATMO (dot-dashed red) and 1D LHD (dotted green) models. The line forming region is approximated by the grey vertical lines which represent the minimum and maximum $\tau_{\rm{R}}( \tau_{\nu} = 1 )$ values in the range 3500 {\AA} and 7200 {\AA} according to Fig.~\ref{fig:tau_ross_comp}. The black vertical lines represent the line forming region if the wavelengths that are within 0.5 {\AA} of
    the line cores are ignored. The bottom boundaries do not change under this definition and therefore overlap with the grey lines.}
    \label{fig:temp_strat_lhd_atmo_threeD}
\end{figure*}

\begin{figure*}
	\includegraphics[width=2\columnwidth]{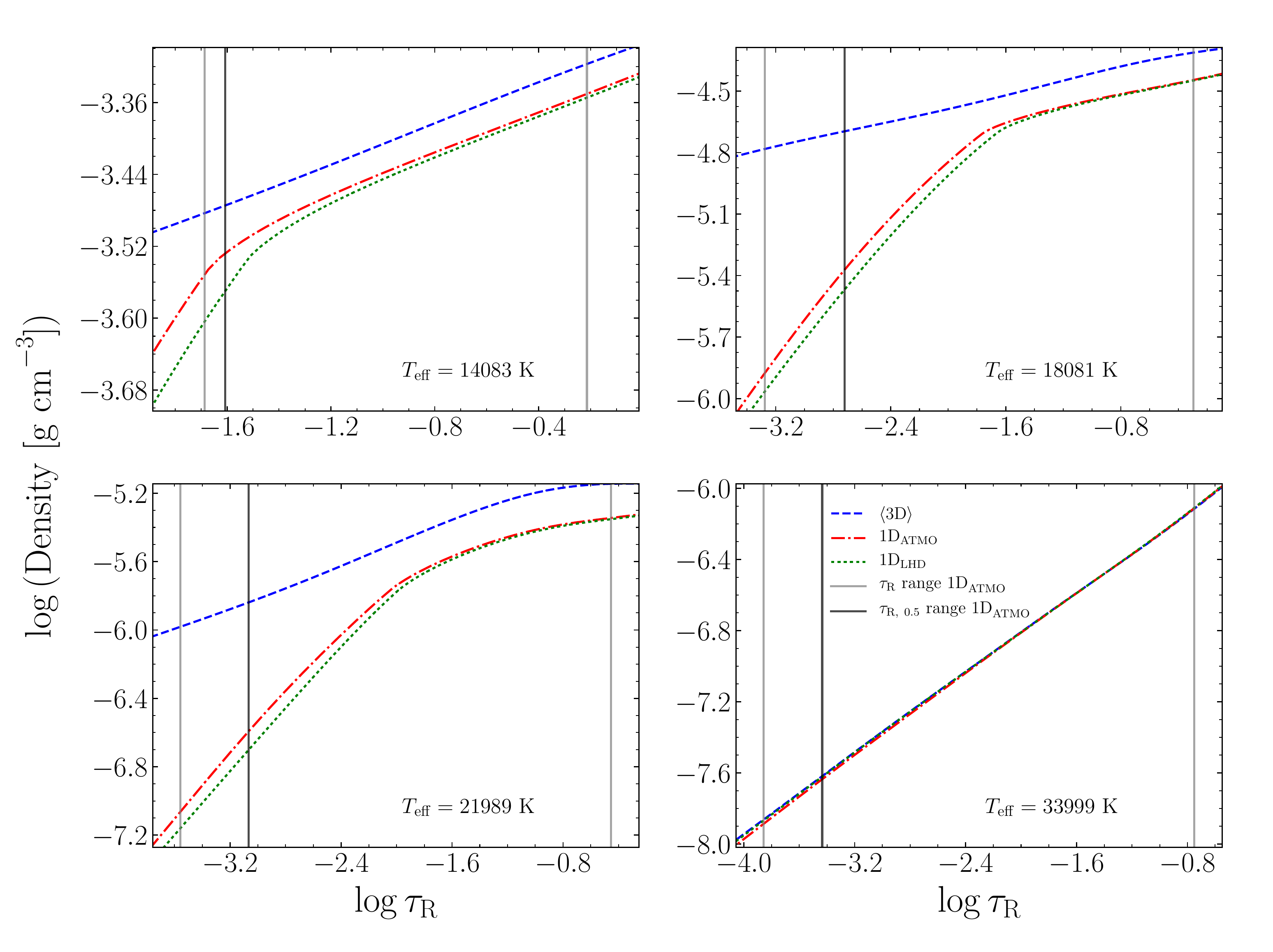}
    \caption{Similar to Fig.~\ref{fig:temp_strat_lhd_atmo_threeD} but for density stratifications of spectral line forming regions for $\langle \rm{3D} \rangle$, 1D ATMO and 1D LHD models.}
    \label{fig:dens_strat_lhd_atmo_threeD}
\end{figure*}

Fig.~\ref{fig:tau_ross_comp} allows for the identification of the atmospheric layers where the continuum and lines between 3500 {\AA} and 7200 {\AA} are formed, so that a comparison can be made between ATMO, LHD and $\langle \rm{3D} \rangle$ structures in the regions relevant to our spectral study. Such a comparison is shown in Figs.~\ref{fig:temp_strat_lhd_atmo_threeD} and \ref{fig:dens_strat_lhd_atmo_threeD} for four of the $\log{g} = 8.0$ models in terms of the temperature and density stratifications, respectively. On these figures we also indicate how the line forming region changes if the wavelengths that are closer than 0.5 {\AA} from the line cores are not included. As the line opacity increases significantly for the line cores, their removal causes the upper boundary of the line forming region to be significantly lower in the atmosphere. This boundary is more appropriate when models are used to fit typical low and medium resolution observations.

The first observation from Figs.~\ref{fig:temp_strat_lhd_atmo_threeD} and \ref{fig:dens_strat_lhd_atmo_threeD} is that for $T_{\rm eff} \sim$ 14\,000, 18\,000, and 22\,000~K, the differences between the $\langle \rm{3D} \rangle$ structures and their 1D counterparts are larger than the differences between 1D ATMO and LHD structures, i.e. the 3D corrections are more significant than the issues with microphysics. Nevertheless, there is some dissension between ATMO and LHD models in this regime, especially at optical depths smaller than the inflexion point above which convection is abruptly switched off as per the prescription of the 1D mixing-length approximation (e.g. $\log{\tau_{\rm{R}}} \lesssim -1.7 $ for the 18\,000~K model). By calculating 1D ATMO structures with scattering treated as true absorption, we found that scattering only has a minor effect in the line forming region and does not significantly improve the agreement between ATMO and LHD. Therefore, we are left with opacity binning as the culprit for the small observed differences between 1D structures at cool temperatures.

At large temperatures (e.g. bottom right plot of Fig.~\ref{fig:temp_strat_lhd_atmo_threeD} at $T_{\rm eff} \sim 34\,000$~K), the disagreement between LHD and ATMO structures becomes more severe in the line forming layers. Interestingly, the good agreement between LHD and $\langle \rm{3D} \rangle$ structures demonstrates that 3D effects are expected to be small at these temperatures. We made attempts to improve the opacity binning procedure or include more bins in LHD (see Section~\ref{sec31}) but could not reach a significantly better agreement. Since the LHD and ATMO codes largely agree at cool temperatures and LHD converges to the 3D simulations in the warm radiative regime, we conclude that it is best to use 1D LHD structures to derive 3D corrections from CO$^5$BOLD simulations. One advantage of this differential analysis is the minimization of the uncertainties caused by the approximations in the microphysics discussed in this section. Furthermore, 3D corrections are generally used for DA white dwarfs rather than the 3D models being used for actual fitting \citep{tremblay_2013_spectra}, suggesting that a differential approach is also advisable for DB white dwarfs. With these justifications, we proceed with 1D LHD structures in the following.

\subsection{3D effects on atmospheric structures}\label{sec23}

\begin{figure}
	\includegraphics[width=\columnwidth]{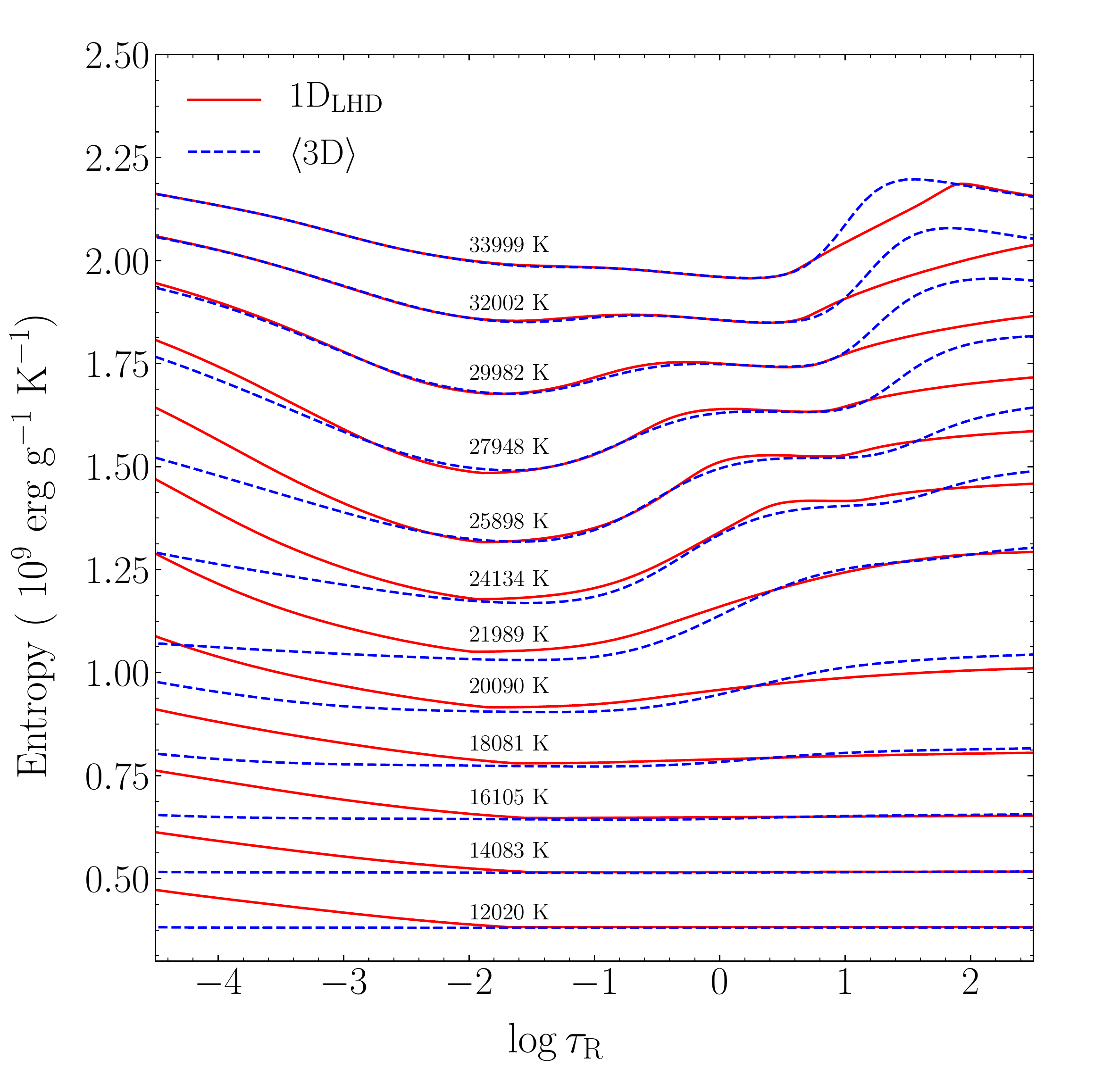}
    \caption{Entropy stratifications for $\log{g} = 8.0$ 1D LHD (solid red) and $\langle \rm{3D} \rangle$ (dashed blue) models identified in Table~\ref{tab:3d_models}. All structures, apart from $T_{\rm{eff}} = 12020$ K, are offset from each other by $0.1 \times 10^9$ erg g$^{-1}$ K$^{-1}$ for clarity.}
    \label{fig:entropy_strat_logg80}
\end{figure}

\begin{figure}
	\includegraphics[width=\columnwidth]{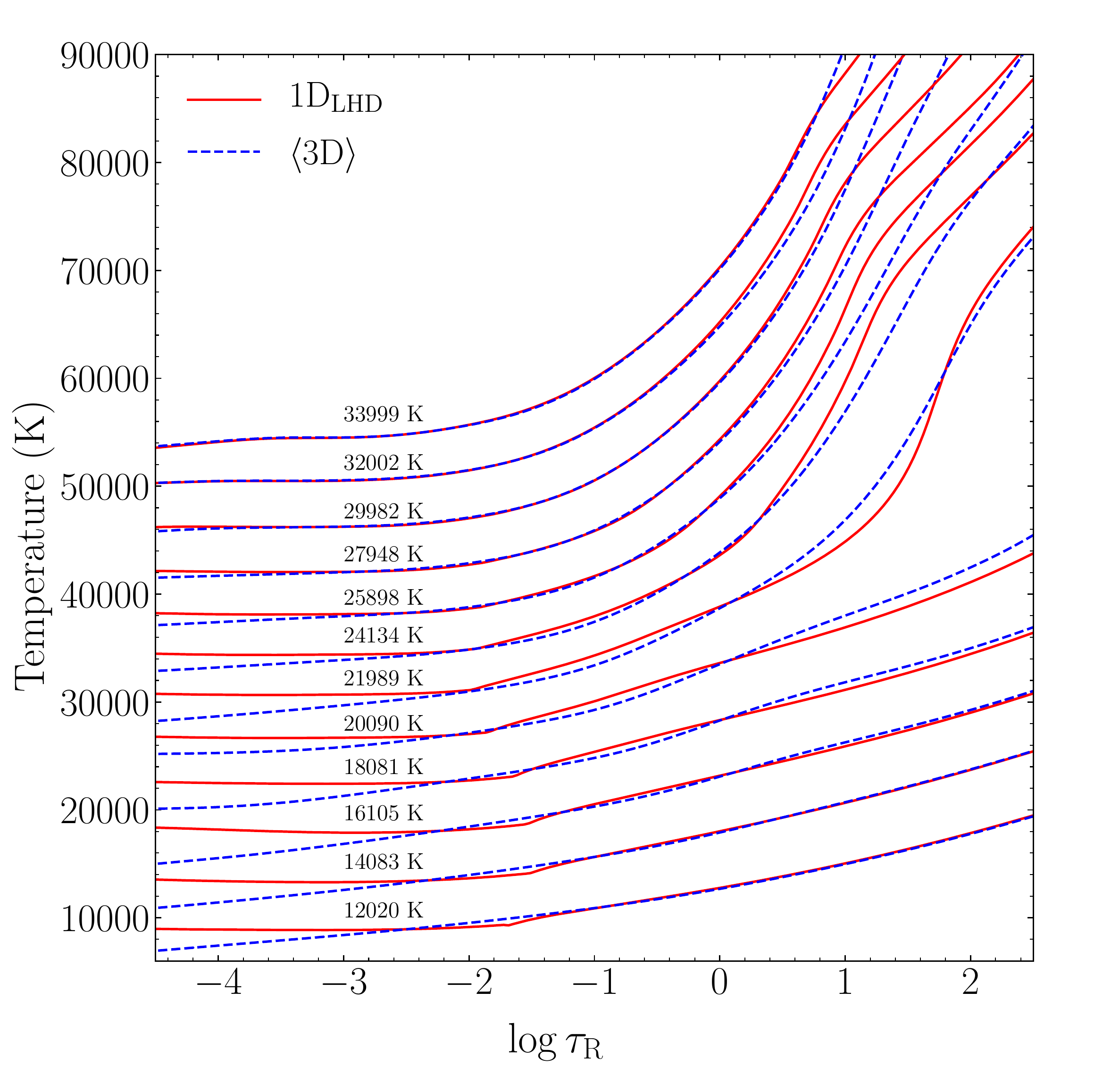}
    \caption{Temperature stratifications for $\log{g} = 8.0$ 1D LHD (solid red) and $\langle \rm{3D} \rangle$ (dashed blue) models identified in Table~\ref{tab:3d_models}. All structures, apart from $T_{\rm{eff}} = 12020$ K, are offset from each other by 3000~K for clarity.}
    \label{fig:temp_strat_lhd_3d_logg80}
\end{figure}

To better understand the structural differences between 1D (LHD) and $\langle \rm{3D} \rangle$ models, Figs.~\ref{fig:entropy_strat_logg80} and~\ref{fig:temp_strat_lhd_3d_logg80} compare the entropy and temperature stratifications for all $\log{g} = 8.0$ models of our 3D grid. Positive and negative entropy gradients as a function of $\tau_{\rm{R}}$ are indicative of atmospheric layers which are unstable and stable against convection, respectively.

For the lowest effective temperatures, good agreement is observed between 1D and $\langle \rm{3D} \rangle$ structures deeper than $\log{\tau_{\rm{R}}} = -1.6$, where convection is adiabatic and therefore both 1D and $\langle \rm{3D} \rangle$ structures converge to the adiabatic gradient. Above these layers, however, the mixing-length approximation predicts no convection and the radiative equilibrium is reached. In the $\langle \rm{3D} \rangle$ picture, overshoot contributes in cooling the upper layers and forces them to have an adiabatic stratification. Very similar results were found for cool DA white dwarfs \citep{tremblay_2013_3dmodels,tremblay_2013_spectra}. Above $T_{\rm{eff}} \sim 16\,000$~K, convection becomes non-adiabatic and sensitive to the prescription of the convective efficiency, resulting in emerging differences between 1D and $\langle \rm{3D} \rangle$ models within the convection zone.

For $T_{\rm{eff}}$ above 24\,000~K but below 34\,000~K, two convective zones develop, associated with He~I and He~II ionization. This is observed for both 1D and $\langle \rm{3D} \rangle$ structures, though for non-local 3D convection, the two convection zones are dynamically connected, and the convective flux remains large in-between the two regions. In this regime the atmospheric structures of DB white dwarfs become more complex compared to DA stars. Convection is driven both by deep optically-thick He~II convection and superficial optically-thin He~I convection, with a thermally stable but dynamically active photosphere in between.  For this DB temperature regime, Fig.~\ref{fig:temp_strat_lhd_3d_logg80} also shows that 3D effects become very small in the line forming layers owing to increasingly inefficient photospheric convection ($\tau_{R} < 1$). Figs.~\ref{fig:entropy_strat_logg80} and~\ref{fig:temp_strat_lhd_3d_logg80} do however suggest strong 3D effects near the bottom of the convection zone for warm simulations, which is related to 1D ML2/$\alpha$ = 1.25 models and 3D simulations predicting significantly different convection zone sizes. We note that this may not be limited to warmer simulations since we do not have access to the bottom of the convection zones for cooler models. The 1D models systematically overpredict the sizes of the convection zones, suggesting that a smaller mixing-length is necessary to match the deep 3D convection zones. We will report on the mixing-length calibration for 1D structures in a future work. This has little to do with the mixing-length value that would be needed for the 1D models to match 3D structures in the line forming regions, which appears neither to be overestimated or underestimated according to Figs.~\ref{fig:entropy_strat_logg80} and~\ref{fig:temp_strat_lhd_3d_logg80}.

\begin{figure*}
    \centering
    \begin{subfigure}[b]{\columnwidth}
        \includegraphics[trim={2cm 1cm 4cm 0cm}, clip, width=\columnwidth]{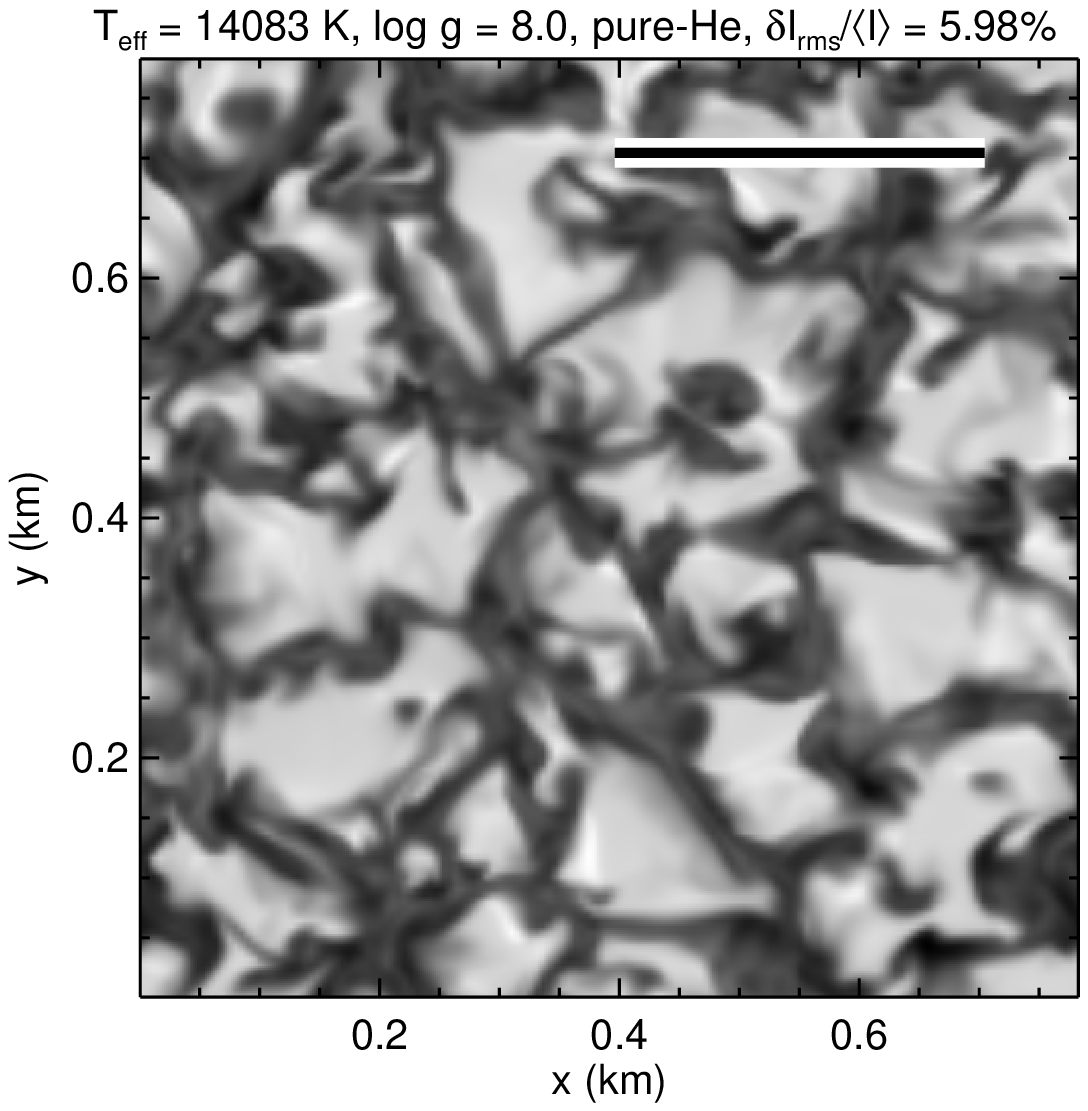}
        \label{fig:d3t140g80db02_xy}
    \end{subfigure}
    ~
    \begin{subfigure}[b]{\columnwidth}
        \includegraphics[trim={2cm 1cm 4cm 0cm}, clip, width=\columnwidth]{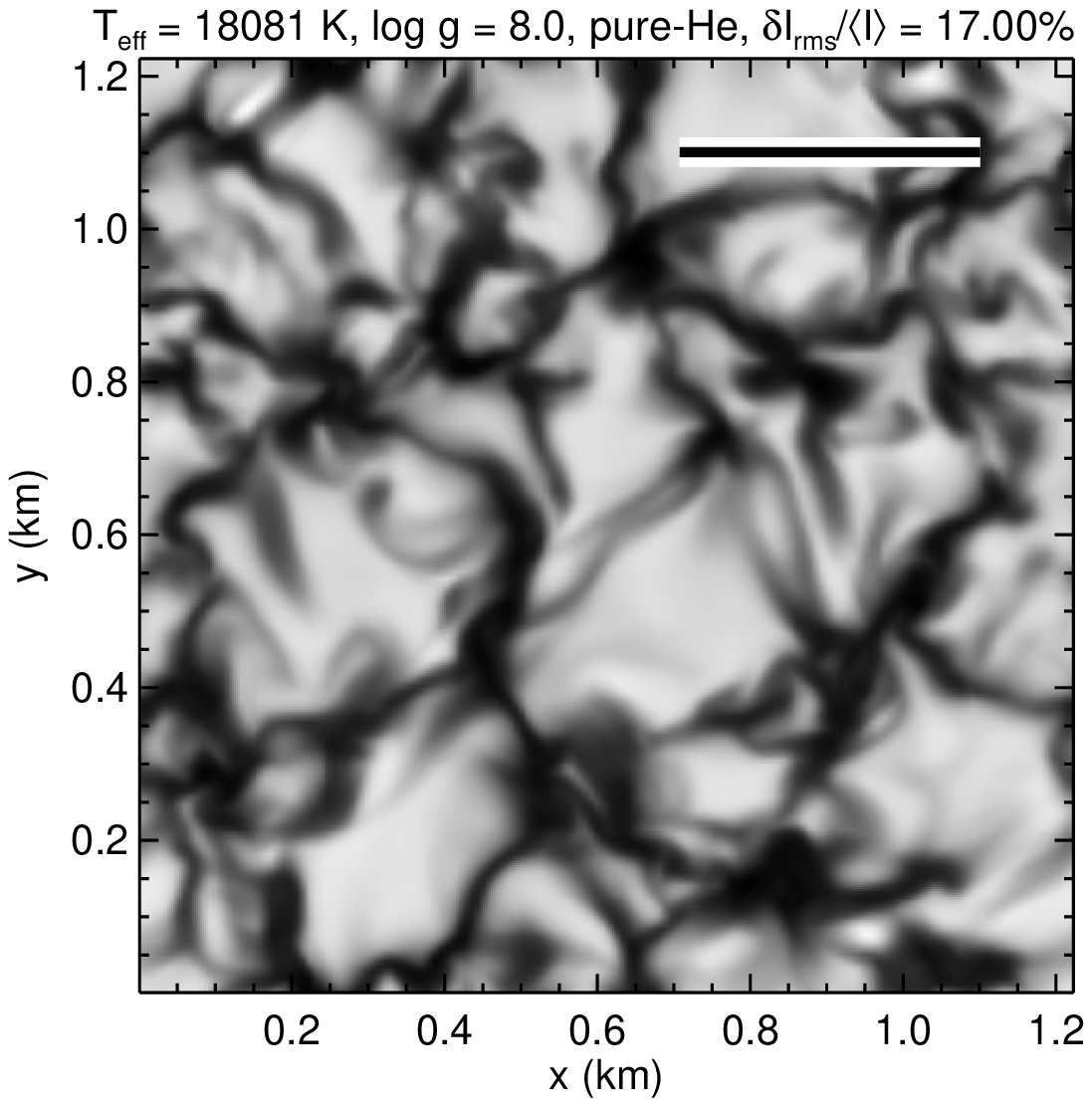}
        \label{fig:d3t180g80db02_xy}
    \end{subfigure}

    \begin{subfigure}[b]{\columnwidth}
        \includegraphics[trim={2cm 1cm 4cm 0cm}, clip, width=\columnwidth]{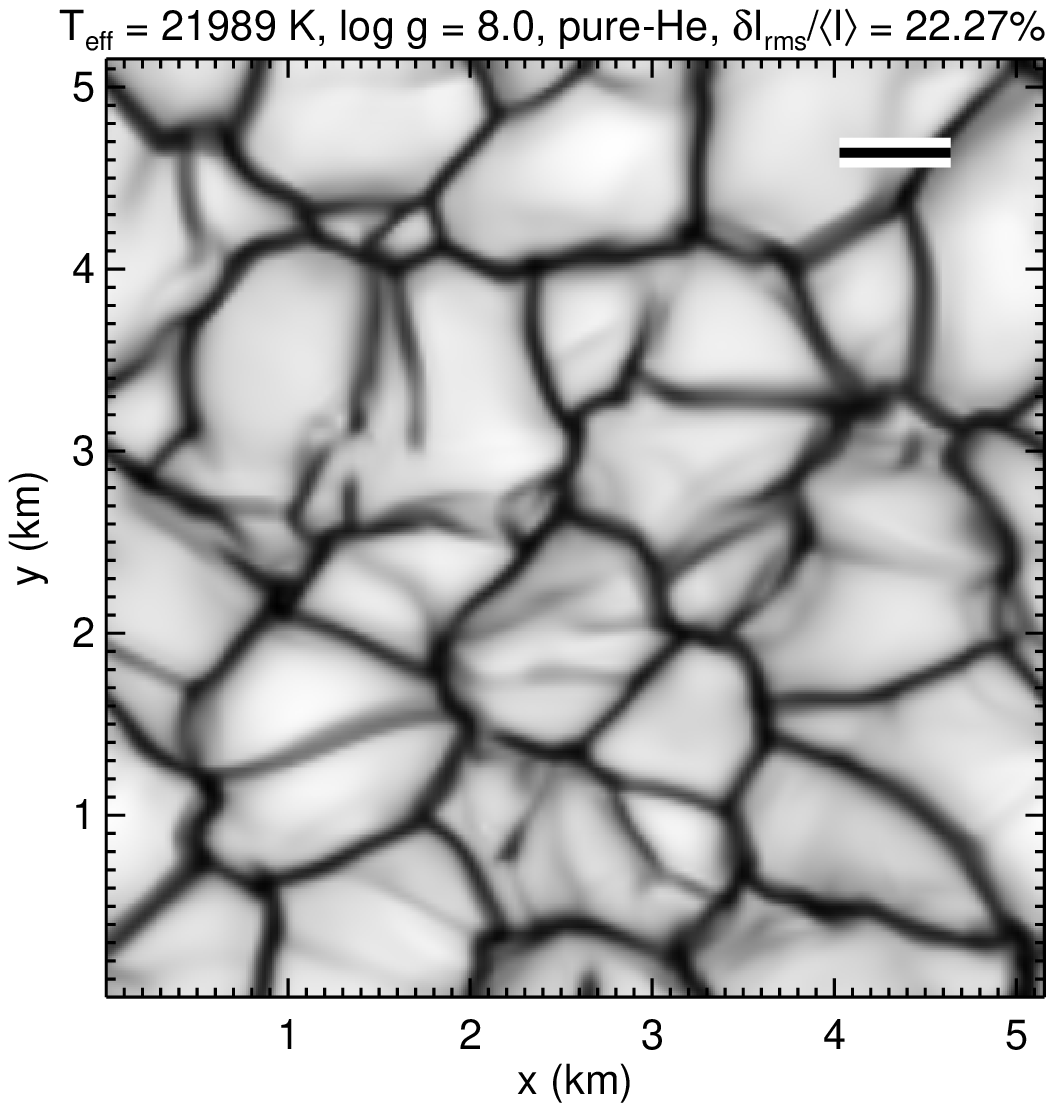}
        \label{fig:d3t220g80db02_xy}
    \end{subfigure}
    ~
    \begin{subfigure}[b]{\columnwidth}
        \includegraphics[trim={2cm 1cm 4cm 0cm}, clip, width=\columnwidth]{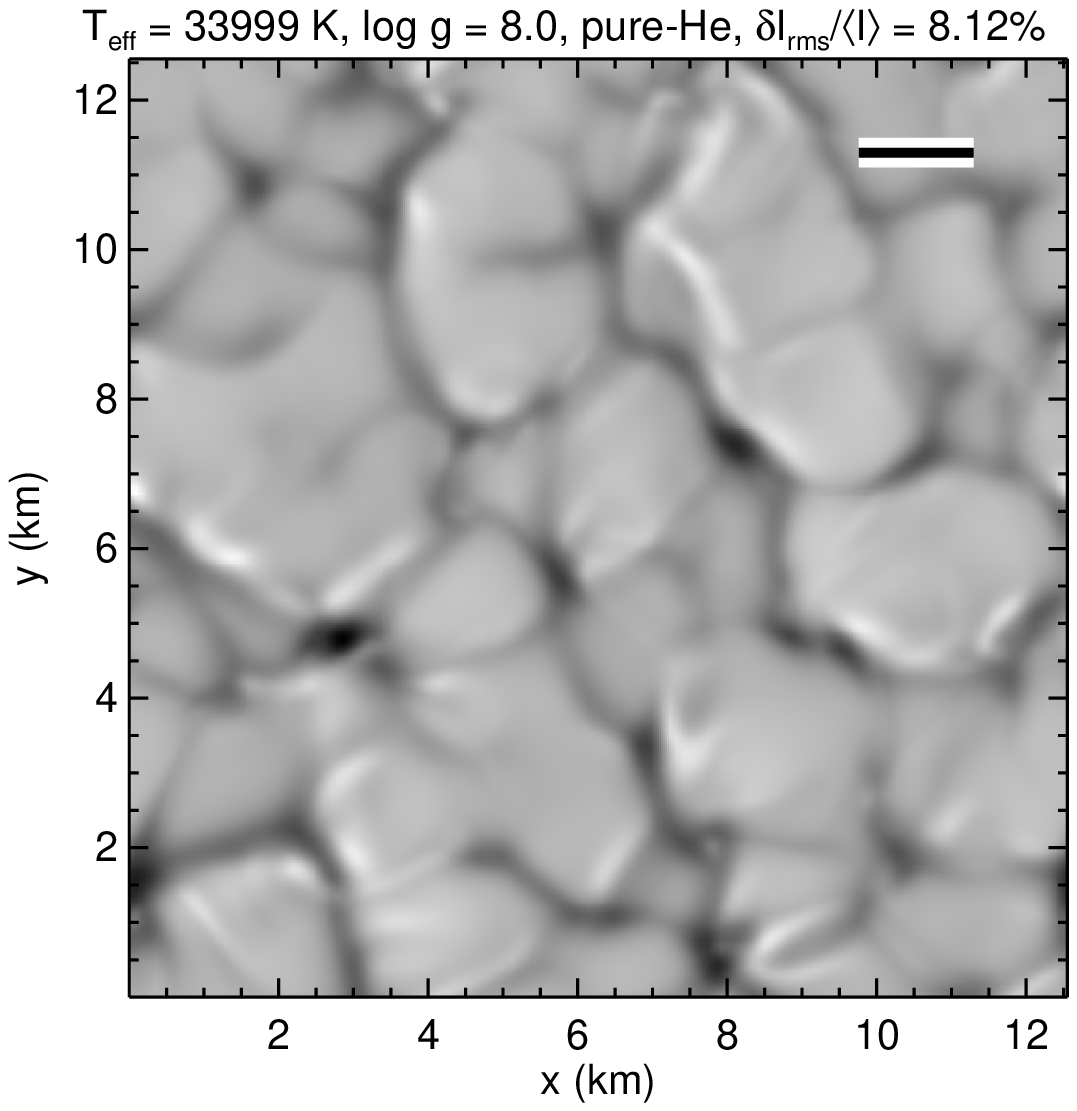}
        \label{fig:d3t340g80db02_xy}
    \end{subfigure}
    \caption{Bolometric emergent intensity for selected models at $\log{g} = 8.0$. In the legends, $T_{\rm eff}$, $\log g$ and intensity contrasts of the simulations are shown. The length of the bar on the top right of each panel is ten times the pressure scale height at $\langle\tau_{\rm{R}}\rangle = 1$.}\label{fig:granulation_plots}
\end{figure*}

Differences between 1D and $\langle \rm{3D} \rangle$ structures can also be understood by looking at the resolved 3D simulations. Fig.~\ref{fig:granulation_plots} shows the bolometric intensity emerging at the top of the simulations for four of the 3D $\log{g} = 8.0$ models. The results are very similar at other surface gravities albeit with a shift in temperature. At low effective temperatures where adiabatic convection dominates, the boundaries of the granules are ill-defined. In this regime the lack of energy loss and the large densities make it possible for convection to transport the required stellar flux with a very small contrast.  For larger effective temperatures convection becomes non-adiabatic and the intensity contrast increases. The radiative timescale decreases such that only the largest granules survive, resulting in a granulation pattern of large cells and narrow intergranular lanes. At $T_{\rm{eff}} \sim 22\,000$~K, the surface of a DB star looks remarkably similar to a DA white dwarf at $T_{\rm{eff}} \sim 12\,000$~K \citep[see Fig.~5 of][]{tremblay_2013_3dmodels}. At $T_{\rm{eff}} \sim 34\,000$ K, convection is very inefficient in the photosphere and the contrast between the cells and intergranular lanes decreases. 

\begin{figure}
	\includegraphics[width=\columnwidth]{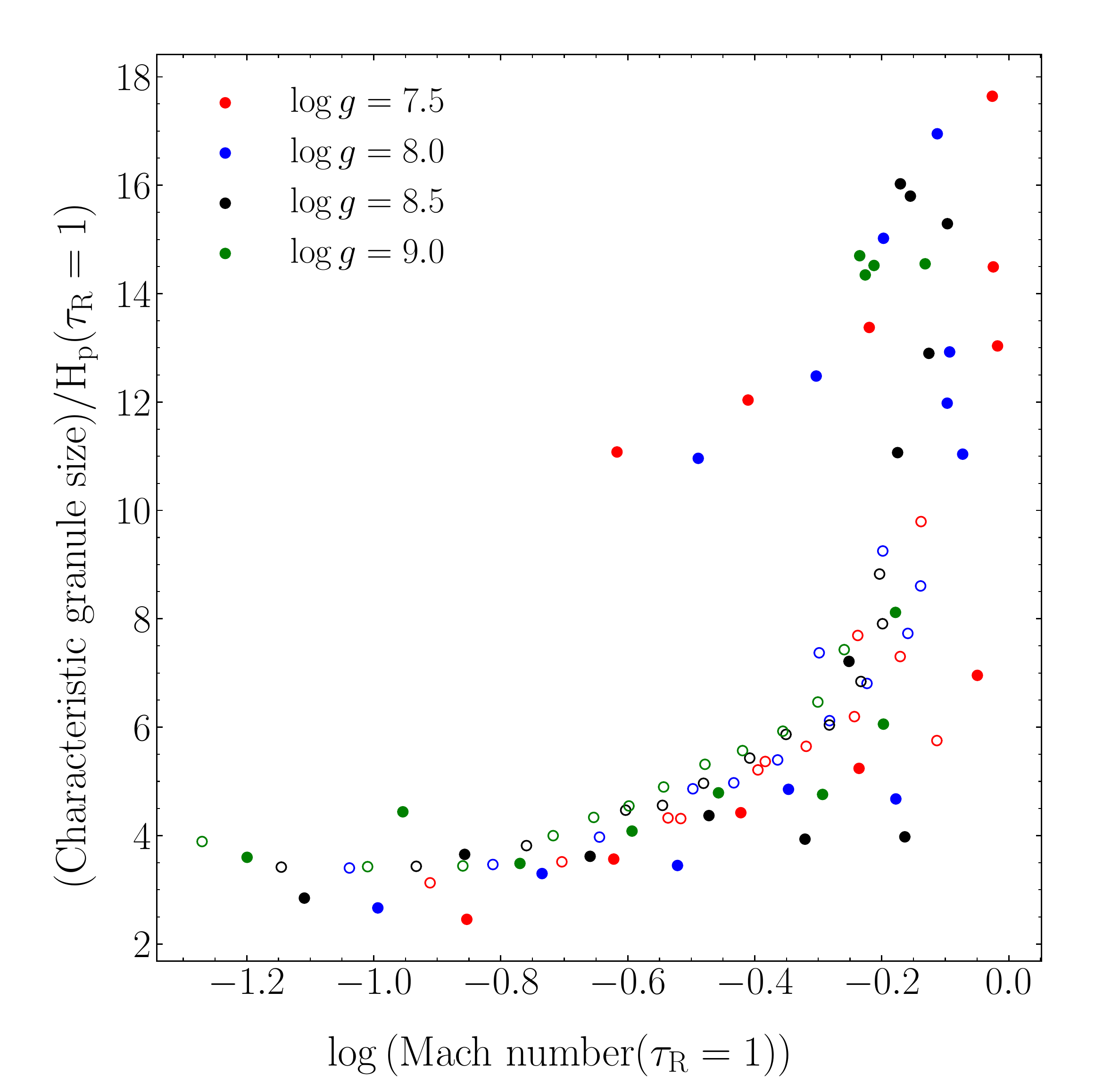}
    \caption{The ratio of the characteristic granule size to the pressure scale height at $\langle\tau_{\rm{R}}\rangle = 1$ as a function of logarithm of the Mach number at $\langle\tau_{\rm{R}}\rangle = 1$ for 3D DB (filled circles) and 3D DA (unfilled circles) models.}
    \label{fig:char_gran_mach}
\end{figure}

In Fig.~\ref{fig:char_gran_mach} we show the ratio of the characteristic granule size to the pressure scale height at $\langle\tau_{\rm{R}}\rangle = 1$ for 3D DA and DB models. In this section all quantities are averaged over constant geometrical depth. The characteristic granule sizes were calculated from the peaks of the emergent intensity power spectra~\citep{tremblay_2013_granulation}. Hotter DB models have granule sizes that are more than ten times the local pressure scale height. Almost all of the models with two convection zones have ratios above 10, suggesting that the presence of He~II convection zone is connected to this behaviour unique to DB white dwarfs. Following the procedure laid out in~\cite{tremblay_2013_granulation} we confirm that the sizes of the granules are consistent with conservation of mass flux and significantly larger horizontal to vertical velocity ratios for hot DB stars.

\begin{figure}
	\includegraphics[width=\columnwidth]{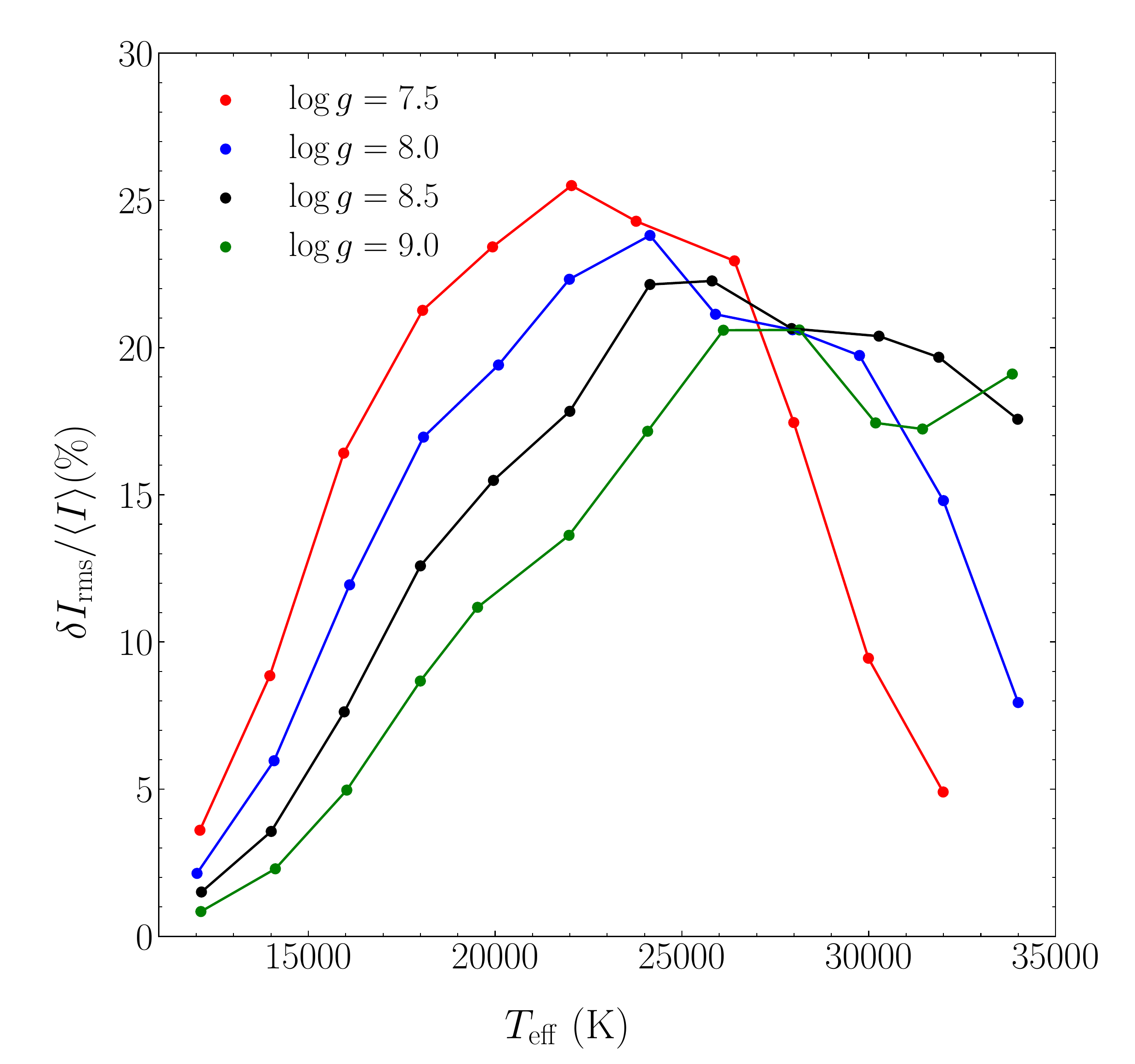}
    \caption{Bolometric intensity contrast as a function of $T_{\rm{eff}}$ for our pure-helium 3D model atmospheres. The points representing intensity contrasts for the same $\log{g}$ values are connected for clarity.}
    \label{fig:i_rms_i_againts_t_plot}
\end{figure}

\begin{figure}
	\includegraphics[width=\columnwidth]{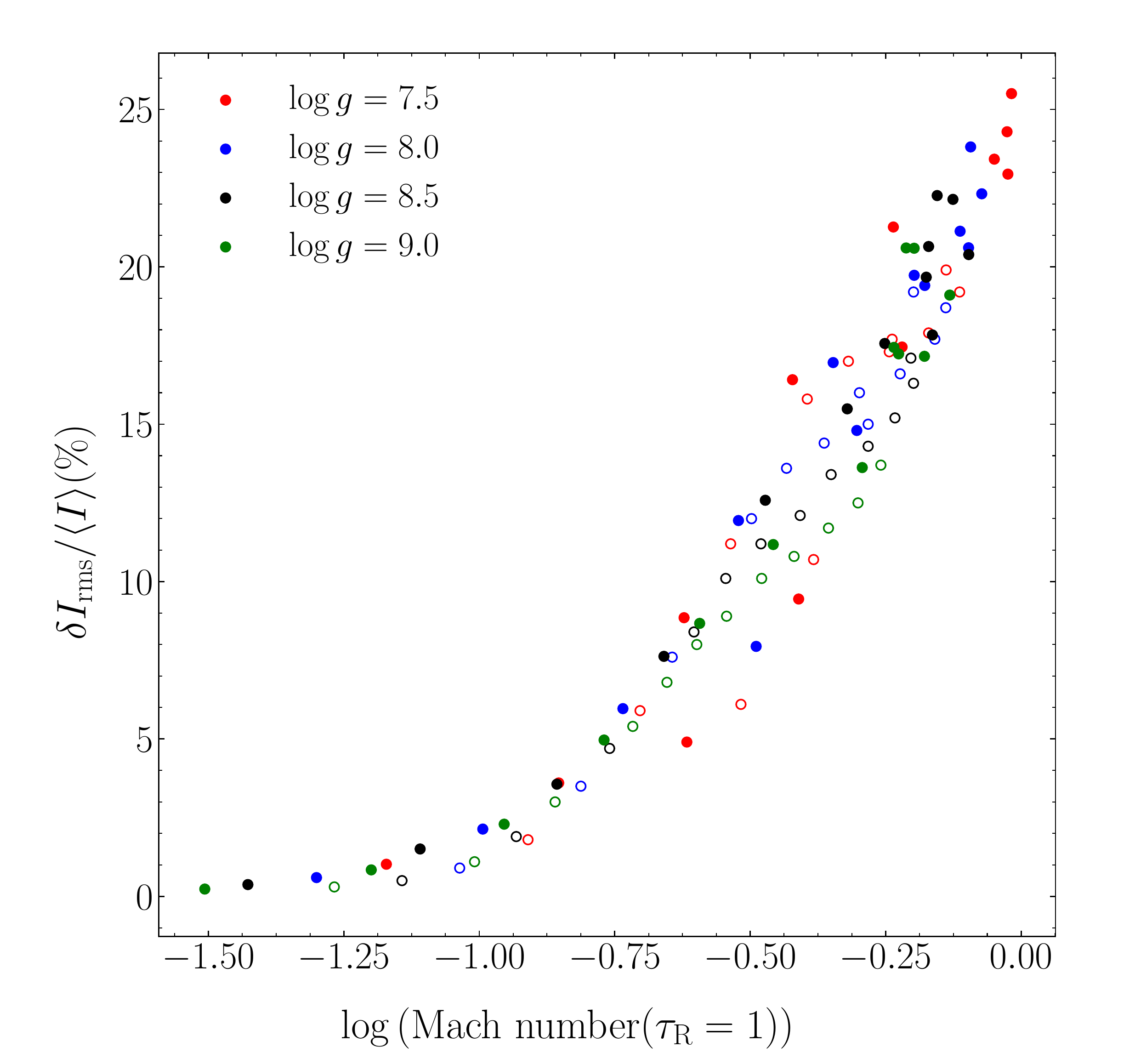}
    \caption{Bolometric intensity contrast as a function of the logarithm of the Mach number at $\langle\tau_{\rm{R}}\rangle = 1$ for our pure-helium 3D model atmospheres (filled circles) and the 3D DA atmospheres of \protect\citet[][open circles]{tremblay_2013_granulation}.}
    \label{fig:mach}
\end{figure}

Figs.~\ref{fig:i_rms_i_againts_t_plot} and \ref{fig:mach} show the intensity contrast as a function of effective temperature and as a function of the Mach number at $\langle\tau_{\rm{R}}\rangle = 1$, respectively. The latter plot also includes data for 3D DA atmospheres from \cite{tremblay_2013_granulation}. We define the Mach number as
\begin{equation}
\text{Mach} = \frac{v_{\rm{rms}}}{c_{\rm{sound}}} = \sqrt{\frac{\langle \rho \rangle v_{\rm{rms}}^2 }{ \langle \Gamma_1 \rangle \langle P \rangle}}~,
\end{equation}
where $v_{\rm{rms}}$ is the convective velocity, $c_{\rm{sound}}$ is the sound speed; $\langle \rho \rangle$, $\langle P \rangle$ and $\langle \Gamma_1 \rangle$ are the geometric horizontal averages of density, pressure and the first adiabatic constant, respectively. Following \cite{tremblay_2013_granulation} $v_{\rm{rms}}^2$ is
\begin{equation}
v_{\rm{rms}}^2 = \langle v^2 \rangle - \frac{[\langle \rho v_x \rangle^2 + \langle \rho v_y \rangle^2 + \langle \rho v_z \rangle^2 ]}{\langle \rho \rangle^2}~,
\end{equation}
where $\langle v^2 \rangle$ is the horizontally averaged mean square velocity and $\langle \rho v_x \rangle^2$, $\langle \rho v_y \rangle^2$, $\langle \rho v_z \rangle^2$ are the three horizontally averaged mass fluxes. The density weighted mean velocity is removed due to its sensitivity to numerical parameters and oscillations.

Both the intensity contrast and the Mach number are measures of the strength of convection and they span a similar range in DA and DB white dwarfs. Helium-atmosphere simulations reach a maximum intensity contrast of about 25\% compared to 20\% for hydrogen-rich compositions. We note that the range in the former case is closer to that seen in main-sequence stars where He ionization is also of relevance \citep{tremblay_2013_granulation}. For a given intensity contrast or Mach number, the density is significantly higher for a DB white dwarf compared to any other convective star, owing to the smaller internal energy density per gram and the larger energy flux to transport. The peak in intensity contrast for DB $\log{g} = 8.0$ models is observed at the effective temperature of 24\,000~K, and above this temperature it significantly decreases and tends towards low values for the $\log{g} = 8.0$ and $\log{g} = 7.5$ models. This is expected for models that are becoming fully radiative. Although it seems as if the intensity contrast is useful to measure the strength of 3D effects on spectra, the link between 3D inhomogeneities and opacities and thus predicted spectral lines is highly non-linear. Furthermore, the strength of 3D effects on spectra also depends on how the different regions of the surface average.

The mean Mach number for a handful of 3D DB models approaches a value of one at the photosphere, indicating that the flows are close to being supersonic. As such, shocks can occur in the simulation and could imprint themselves on synthetic spectra. We note that the situation is no different to DA white dwarfs or main-sequence stars for which the mean Mach number can reach a value close to one.

\section{Model spectra}\label{sec:model_spectra}

Neither LHD nor CO$^{5}$BOLD can perform detailed spectral synthesis. Given our differential approach at comparing 1D LHD and $\langle \rm{3D} \rangle$ structures, it is thus appropriate to use the ATMO code of BW11 to calculate synthetic spectra. We employ the same numerical setup as that used by BW11 to compute their DB grid.

To calculate differential 3D corrections, we use the DB fitting code of BW11 to fit $\langle$3D$\rangle$ synthetic spectra with the 1D (LHD) spectral grid. This ensures that we consider the same wavelength region and the same lines BW11 analysed in their study. We define the 3D corrections to be
\begin{equation}
T_{\rm{eff, \ corr}} = T_{\rm{eff, \ 3D \ model}} - T_{\rm{eff, \ 1D \ fit}}~,
\label{eq2}
\end{equation}
and
\begin{equation}
\log{g}_{\rm{corr}} = \log{g}_{\rm{3D \ model}} - \log{g}_{\rm{1D \ fit}}~.
\label{eq3}
\end{equation}
Since DBA stars were also included in their study, BW11 used the code to either fit hydrogen lines or to apply upper limits in the case of non-detection. As our models are pure-helium, we have instead adapted the code to have pure-helium composition as the only option. Before presenting our proposed 3D corrections, we first evaluate the uncertainties from the different approximations we have made, namely the opacity binning procedure and the mean 3D approximation.

\subsection{Effect of opacity binning}\label{sec31}

\begin{figure*}
	\includegraphics[width=2\columnwidth]{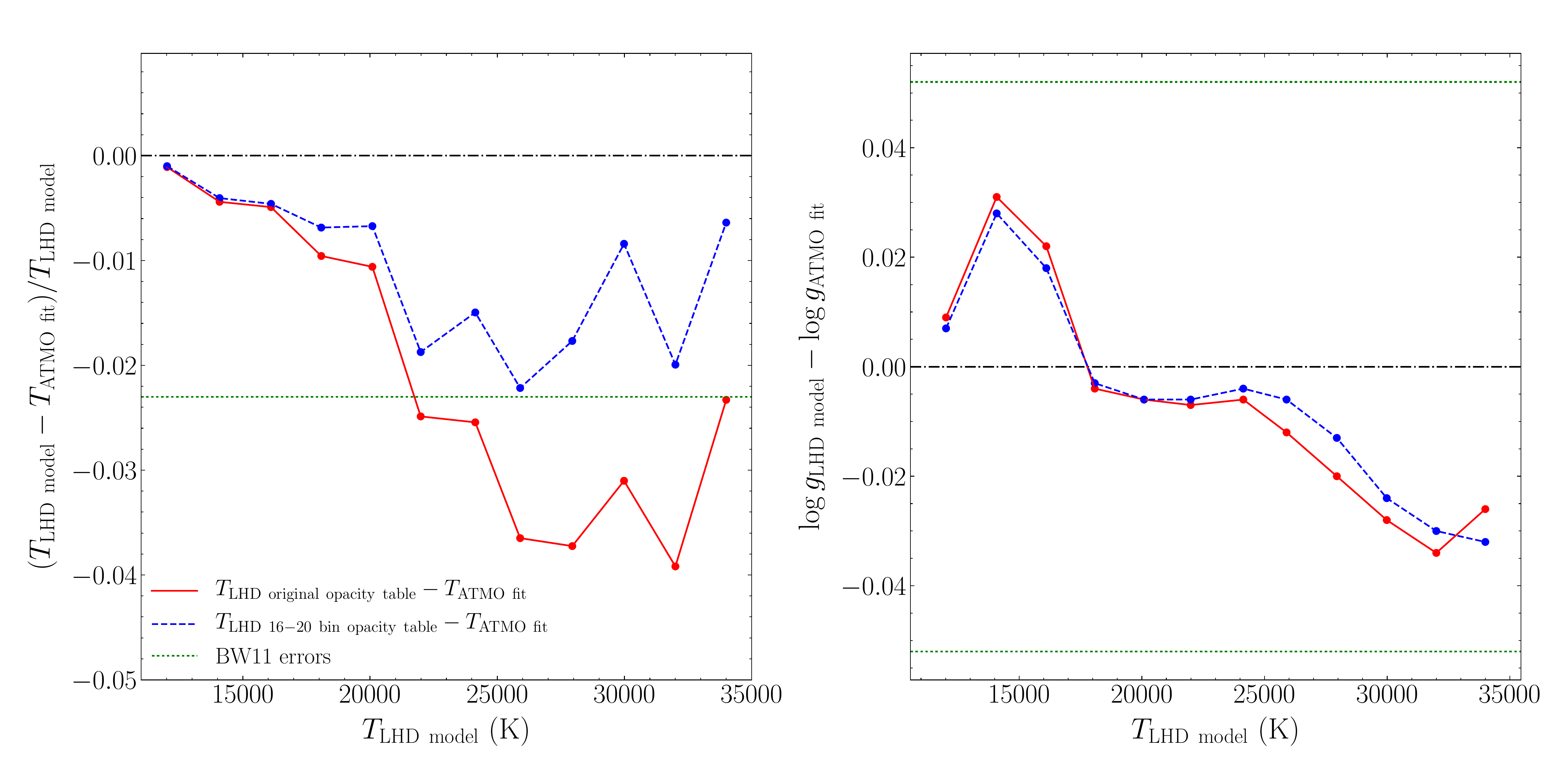}
    \caption{Fits of synthetic spectra based on 1D LHD $\log{g} = 8.0$ structures that have been computed with two types of opacity tables: original (10 bins, solid red) and extended (16-20 bins, dashed blue). These model spectra were fitted with a grid of standard 1D ATMO spectra of BW11 to quantify the differences between the two types of 1D codes. The resulting $T_{\rm{eff}}$ and $\log g$ corrections are presented on the left and right panels, respectively. The uncertainties quoted by BW11 found from comparing multiple spectra of 28 DB and DBA white dwarfs are also shown with dotted green lines. A dot-dashed horizontal black line representing a null correction has also been added to each panel for clarity.}
    \label{fig:lhd_atmo_corr_logg80}
\end{figure*}

Two sets of $\log{g} = 8.0$ 1D LHD structures were computed with the same effective temperatures as the 3D models. The first set was computed with 10 bin opacity tables already employed for our 3D models and we shall refer to these as ``LHD$_{\rm original}$''. The other ``LHD$_{\rm 16-20 bins}$'' set was calculated using opacity tables with 16 to 20 bins, which do not remove the large far-UV opacity unlike the 10 bin opacity tables. We have derived synthetic spectra using ATMO for the two sets of LHD structures. The full grid of BW11 1D DB synthetic spectra is also used for opacity binning analysis. This grid was calculated from 1D structures computed with the ATMO code and therefore all 1745 frequencies were used in the computation instead of opacity binning. To quantify the corrections arising from the opacity binning, the two types of LHD spectra were fitted with the 1D ATMO spectral grid and the differences between the atmospheric parameters are shown in Fig.~\ref{fig:lhd_atmo_corr_logg80}. A negative difference indicates that ATMO overestimates the $T_{\rm{eff}}$ or $\log{g}$ of the LHD spectrum. BW11 determined external errors for their DB and DBA survey by comparing fitted values for multiple spectra of 28 stars. They found average uncertainties of $\langle\Delta{T_{\rm{eff}}}/T_{\rm{eff}}\rangle = 2.3 \%$ and $\langle\Delta{\log{g}}\rangle = 0.052$ when obvious outliers are removed (see their Figure~17). These errors are plotted on Fig.~\ref{fig:lhd_atmo_corr_logg80} and are referred to as BW11 errors. In comparison, \cite{koester_kepler_2015} average external uncertainties for their SDSS sample with lower average signal-to-noise are $\langle\Delta{T_{\rm{eff}}}/T_{\rm{eff}}\rangle = 3.1 \%$ and $\langle\Delta{\log{g}}\rangle = 0.12$.

The $\log{g}$ corrections for both original and extended opacity tables are well within fitting uncertainties (right panel of Fig.~\ref{fig:lhd_atmo_corr_logg80}). For $T_{\rm{eff}}$ corrections only the LHD models with extended opacity tables fall within the uncertainties. As expected from our discussion around Figs.~\ref{fig:temp_strat_lhd_atmo_threeD} and \ref{fig:dens_strat_lhd_atmo_threeD} and the comparison of ATMO and LHD structures, the largest differences are expected for the warmest simulations. Although the agreement improves when doubling the number of opacity bins, we did not pursue the possibility of improving the opacity binning procedure for LHD and by inference our CO$^5$BOLD simulations because of the dramatic increase in computation time. Instead we employ our 10 bin tables, and a differential approach between LHD and CO$^5$BOLD structures which largely removes the offset observed in Fig.~\ref{fig:lhd_atmo_corr_logg80}. 

\subsection{Mean 3D approximation}\label{sec32}

Ideally, 3D spectral synthesis is performed to compute a spectrum directly from a 3D data cube. One such code, Linfor3D \citep{ludwig08}, was utilised by \cite{tremblay11} to model synthetic 3D H$\beta$ lines for DA white dwarfs. While the code could be adapted to synthesise selected 3D spectral lines for DB white dwarfs, it would be computationally expensive to create a full grid of 3D model spectra. Instead, we proceed with the comparison of two types of estimates for calculations of synthetic spectra from 3D simulations: the $\langle$3D$\rangle$ and 1.5D approximations. Our standard $\langle$3D$\rangle$ spectra are computed from the $\langle$3D$\rangle$ temperature and pressure structures using ATMO. On the other hand, the 1.5D method assumes that each 3D simulation, which is made up of $150 \times 150 \times 150$ grid points, is a collection $150 \times 150$ ``1D'' atmospheres, where the vertical extent of the simulation ($z$-axis) is the extent of these 1D atmospheres. For each of the ``1D'' atmospheres a spectrum is then calculated using ATMO and the resulting $150 \times 150$ spectra are simply averaged to produce a so-called 1.5D spectrum for a given 3D model. However, we found that some of the atmospheres exhibited pressure inversion due to the departure from hydrostatic equilibrium, which is expected in 3D simulations. ATMO is not adapted to handle such departures and therefore any structures with pressure inversion were removed from the 1.5D spectrum calculations. We also want the 1.5D spectrum to be representative of the entire simulation and not of one single time snapshot, and therefore we used several snapshots over the last quarter of the computation for the average. The $\langle$3D$\rangle$ and 1.5D methods represent the two extremes in neglecting or enhancing the 3D fluctuations, respectively, and thus the full 3D spectral synthesis is somewhere in between these two methods. For the majority of 3D DA models, with the exception of extremely low-mass (ELM) white dwarfs, it has been shown that 1.5D and $\langle$3D$\rangle$ corrections are equivalent \citep{tremblay15}. This results from a complex cancellation of the 3D fluctuations in spectral synthesis \citep{tremblay_2013_spectra} and there is no obvious reason to assume the same behaviour for DB white dwarfs.

\begin{figure*}
	\includegraphics[width=2\columnwidth]{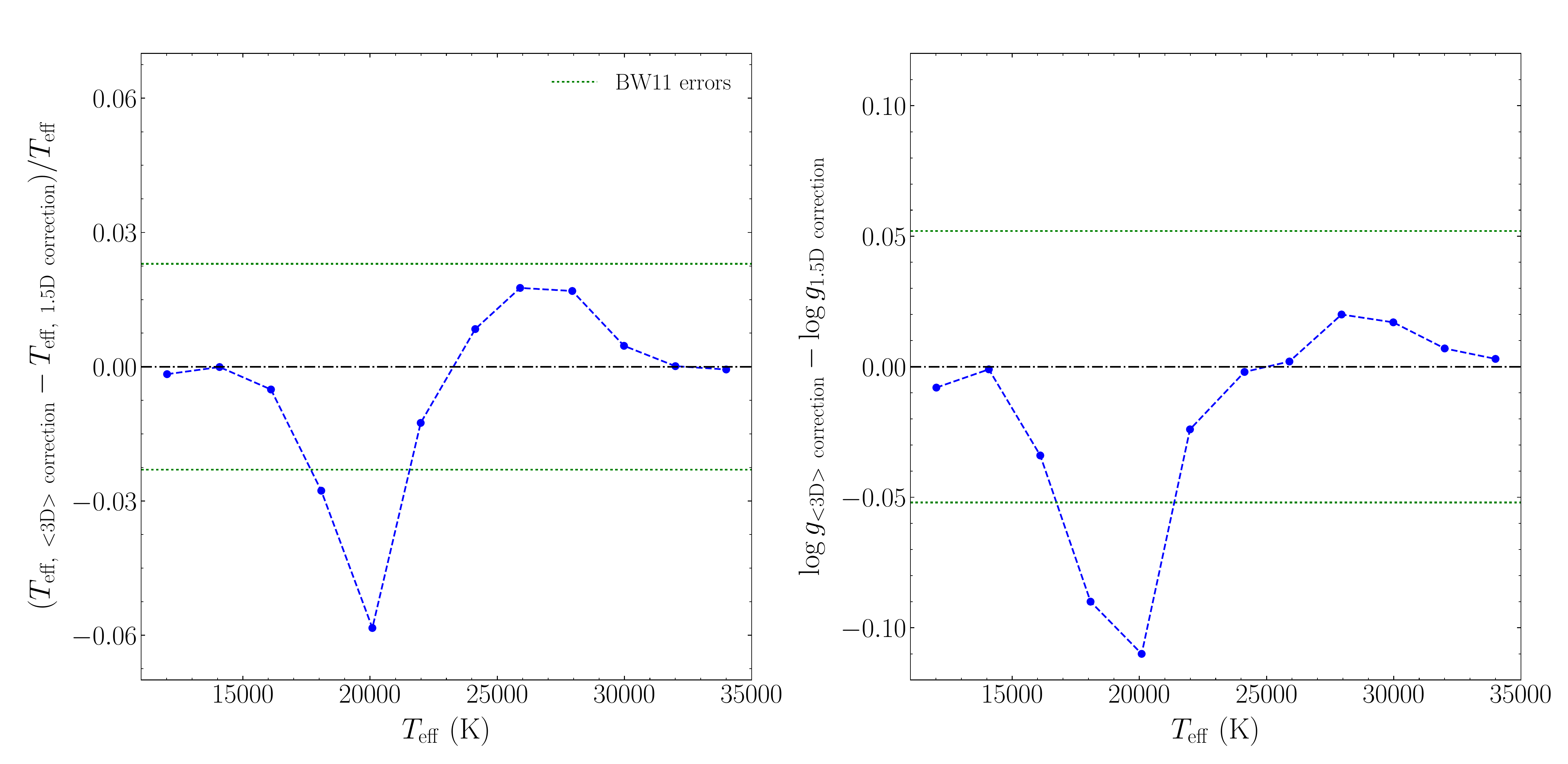}
    \caption{1.5D and $\langle$3D$\rangle$ spectra at $\log{g} = 8.0$ are compared in terms of the differences between their respective $T_{\rm eff}$ (left panel) and $\log g$ (right panel) corrections found by fitting the two types of spectra with the 1D LHD models. External fitting uncertainties from BW11 are also shown. A dot-dashed horizontal black line representing the equivalence of 1.5D and $\langle$3D$\rangle$ corrections (and therefore spectra) has also been added to each panel.}
    \label{fig:lhd_3d_15d_corr_og}
\end{figure*}

To determine the uncertainties arising from not using the full 3D spectral synthesis, we fitted both the 1.5D and $\langle$3D$\rangle$ spectra with the 1D LHD model grid to find their respective corrections. Fig.~\ref{fig:lhd_3d_15d_corr_og} shows the differences between the $\langle$3D$\rangle$ and 1.5D corrections for $\log{g}=8.0$ models, although similar results are obtained for other $\log{g}$ models. The BW11 errors are also shown. A negative difference means that the 1.5D correction is larger and this is what we observe for 3D simulations below $\approx$ 24\,000~K. Most of the differences are well within the BW11 external errors, with the maximum offset observed at $\approx$ 20\,000~K. At this particular effective temperature, He~I lines reach their maximum strength (depending on the assumed convective efficiency) giving rise to the hot/cool solution problem, where for any given DB spectrum there are often two possible fits with equivalent $\chi^2$ values. BW11 have also shown that in the range $20\,000 \lesssim T_{\rm{eff}} \lesssim 28\,000$ K, the spectra are quite insensitive to the effective temperature. This suggests that fitting uncertainties may peak in this region even though we have employed a constant average uncertainty in percentage. 

Another possibility for the disagreement between 1.5D and $\langle$3D$\rangle$ corrections could be related to the high Mach numbers of some of the DB simulations as 1.5D spectra are more sensitive to thermal fluctuations caused by shocks. For $\log{g} = 8.0$ simulations, the $T_{\rm{eff}} \approx 22\,000$~K model has the highest Mach number, and yet for this particular model the $\langle$3D$\rangle$ and 1.5D corrections do agree, suggesting that there is no obvious link.

We stress that since the full 3D spectral synthesis is expected to lie somewhere between the 1.5D and $\langle$3D$\rangle$ corrections, Fig.~\ref{fig:lhd_3d_15d_corr_og} likely overestimates the error of using the $\langle$3D$\rangle$ approximation. We conclude that the $\langle$3D$\rangle$ approximation is valid for DB white dwarfs, a result that is similar to that found for DA stars above $\log g = 7.0$.

\section{Discussion}\label{sect:3d_corr}

\subsection{3D corrections}

\begin{figure}
	\includegraphics[width=\columnwidth]{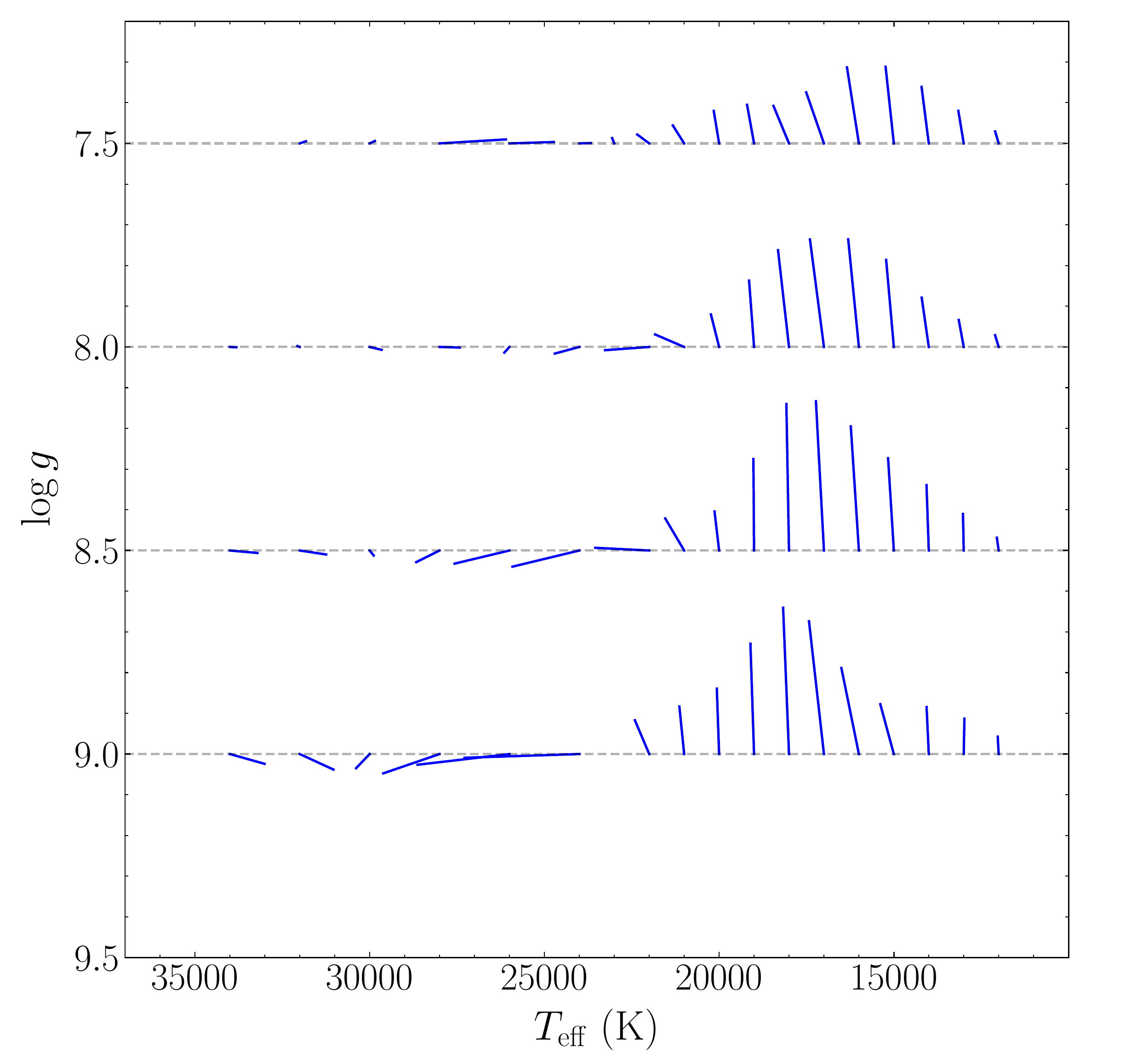}
    \caption{Proposed 3D corrections for pure-helium white dwarf atmospheres when compared with 1D model spectra using the ML2/$\alpha$ = 1.25 parameterisation. The horizontal dashed lines indicate $\log{g} =$~[7.5, 8.0, 8.5, 9.0], whereas the intersections between the dashed lines and the blue lines are the reference 1D atmospheric parameters. If one follows a given blue line from the intersection, its length will give the 3D correction for the particular 1D $\log{g}$ and $T_{\rm{eff}}$. For clarity, at high effective temperatures some of the corrections from Table~\ref{tab:3d_corr} are not shown. This figure can be compared to 3D DA corrections from \protect\cite{tremblay15}.}
    \label{fig:final_3d_corr}
\end{figure}

Our proposed 3D corrections for pure-helium DB white dwarfs are shown in Fig.~\ref{fig:final_3d_corr} and are tabulated in Table~\ref{tab:3d_corr}. The corrections were derived using the reference 1D LHD spectral grid under the ML2/$\alpha$ = 1.25 parameterisation. In the table, the corrections quoted are calculated using Eqs.~\ref{eq2} and \ref{eq3} and are interpolated over for reference 1D $\log{g}$ and $T_{\rm{eff}}$ values. In the figure, the dashed lines denote $\log{g} =$ [7.5, 8.0, 8.5, 9.0]. The intersection points between the dashed horizontal lines and the blue lines are the 1D $\log{g}$ and $T_{\rm{eff}}$ values. The blue lines then extend to the corresponding 3D parameters, such that the lengths of the blue lines represent our proposed 3D corrections for the 1D parameters. The main uncertainty in the 3D corrections resides in the $\langle$3D$\rangle$ approximation discussed in Section~3.2, but also important is the validity of the pure-helium atmosphere approximation when applying the corrections to specific DB white dwarfs.

\begin{table}
	\centering
	\caption{Our proposed 3D corrections for $\log{g}$ and $T_{\rm{eff}}$ derived from $\langle$3D$\rangle$ structures. A negative value indicates that 1D overestimates the parameter, while a positive value indicates underestimation.}
	\label{tab:3d_corr}
	\begin{tabular}{lccccr} 
		\hline
		1D $\log{g}$ & 1D $T_{\rm{eff}}$ (K) & 3D $\log{g}$ & 3D $T_{\rm{eff}}$ \\
		& & correction (dex) & correction (K) \\ 
		\hline        
        7.5 & 12000 & $-$0.029 & 100 \\
        7.5 & 13000 & $-$0.080 & 154 \\
        7.5 & 14000 & $-$0.139 & 206 \\
        7.5 & 15000 & $-$0.188 & 236 \\
        7.5 & 16000 & $-$0.187 & 340 \\
        7.5 & 17000 & $-$0.126 & 507 \\
        7.5 & 18000 & $-$0.092 & 444 \\
        7.5 & 19000 & $-$0.095 & 200 \\
        7.5 & 20000 & $-$0.080 & 153 \\
        7.5 & 21000 & $-$0.044 & 324 \\
        7.5 & 22000 & $-$0.022 & 341 \\
        7.5 & 23000 & $-$0.013 & 62 \\
        7.5 & 24000 & $-$0.001 & $-$328 \\
        7.5 & 25000 & 0.002 & $-$710 \\
        7.5 & 26000 & $-$0.003 & $-$1268 \\
        7.5 & 27000 & $-$0.009 & $-$1915 \\
        7.5 & 28000 & $-$0.010 & $-$1897 \\
        7.5 & 29000 & $-$0.004 & $-$201 \\
        7.5 & 30000 & $-$0.006 & $-$148 \\
        7.5 & 31000 & $-$0.006 & $-$165 \\
        7.5 & 32000 & $-$0.005 & $-$176 \\
        7.5 & 33000 & $-$0.003 & $-$218 \\
        7.5 & 34000 & $-$0.000 & 149 \\
        \hline
        8.0 & 12000 & $-$0.029 & 101 \\
        8.0 & 13000 & $-$0.066 & 142 \\
        8.0 & 14000 & $-$0.121 & 202 \\
        8.0 & 15000 & $-$0.214 & 219 \\
        8.0 & 16000 & $-$0.264 & 306 \\
        8.0 & 17000 & $-$0.264 & 400 \\
        8.0 & 18000 & $-$0.238 & 314 \\
        8.0 & 19000 & $-$0.163 & 142 \\
        8.0 & 20000 & $-$0.080 & 234 \\
        8.0 & 21000 & $-$0.031 & 833 \\
        8.0 & 22000 & 0.008 & 1264 \\
        8.0 & 23000 & 0.015 & 1055 \\
        8.0 & 24000 & 0.016 & 703 \\
        8.0 & 25000 & 0.017 & 558 \\
        8.0 & 26000 & 0.014 & 146 \\
        8.0 & 27000 & 0.011 & $-$484 \\
        8.0 & 28000 & 0.001 & $-$578 \\
        8.0 & 29000 & 0.003 & $-$105 \\
        8.0 & 30000 & 0.007 & $-$342 \\
        8.0 & 31000 & 0.003 & $-$252 \\
        8.0 & 32000 & $-$0.002 & 66 \\
        8.0 & 33000 & $-$0.001 & $-$627 \\
        8.0 & 34000 & 0.001 & $-$186 \\
        \hline
        8.5 & 12000 & $-$0.031 & 51 \\
        8.5 & 13000 & $-$0.090 & 21 \\
        8.5 & 14000 & $-$0.161 & 65 \\
        8.5 & 15000 & $-$0.227 & 164 \\
        8.5 & 16000 & $-$0.305 & 231 \\
        8.5 & 17000 & $-$0.367 & 229 \\
        8.5 & 18000 & $-$0.360 & 75 \\
        8.5 & 19000 & $-$0.225 & 18 \\
        8.5 & 20000 & $-$0.096 & 130 \\
        8.5 & 21000 & $-$0.079 & 541 \\
        8.5 & 22000 & $-$0.007 & 1543 \\
        8.5 & 23000 & 0.032 & 1886 \\
        \hline
	\end{tabular}
\end{table}

\begin{table}
	\centering
	\contcaption{}
	\begin{tabular}{lccccr} 
		\hline
		1D $\log{g}$ & 1D $T_{\rm{eff}}$ (K) & 3D $\log{g}$ & 3D $T_{\rm{eff}}$ \\
		& & correction (dex) & correction (K) \\ 
		\hline     
        8.5 & 24000 & 0.040 & 1913 \\
        8.5 & 25000 & 0.033 & 1915 \\
        8.5 & 26000 & 0.032 & 1570 \\
        8.5 & 27000 & 0.037 & 1068 \\
        8.5 & 28000 & 0.029 & 666 \\
        8.5 & 29000 & 0.021 & 396 \\
        8.5 & 30000 & 0.012 & $-$116 \\
        8.5 & 31000 & 0.004 & $-$455 \\
        8.5 & 32000 & 0.010 & $-$755 \\
        8.5 & 33000 & 0.011 & $-$1580 \\
        8.5 & 34000 & 0.006 & $-$789 \\
        \hline		
        9.0 & 12000 & $-$0.043 & 26 \\
        9.0 & 13000 & $-$0.087 & $-$19 \\
        9.0 & 14000 & $-$0.115 & 62 \\
        9.0 & 15000 & $-$0.123 & 387 \\
        9.0 & 16000 & $-$0.211 & 500 \\
        9.0 & 17000 & $-$0.326 & 427 \\
        9.0 & 18000 & $-$0.360 & 168 \\
        9.0 & 19000 & $-$0.271 & 104 \\
        9.0 & 20000 & $-$0.161 & 65 \\
        9.0 & 21000 & $-$0.117 & 137 \\
        9.0 & 22000 & $-$0.083 & 410 \\
        9.0 & 23000 & $-$0.069 & 1042 \\
        9.0 & 24000 & 0.009 & 3279 \\
        9.0 & 25000 & 0.018 & 3071 \\
        9.0 & 26000 & 0.026 & 2629 \\
        9.0 & 27000 & 0.042 & 2119 \\
        9.0 & 28000 & 0.047 & 1610 \\
        9.0 & 29000 & 0.045 & 1110 \\
        9.0 & 30000 & 0.035 & 389 \\
        9.0 & 31000 & 0.030 & $-$333 \\
        9.0 & 32000 & 0.038 & $-$963 \\
        9.0 & 33000 & 0.040 & $-$1434 \\
        9.0 & 34000 & 0.024 & $-$993 \\
        \hline
	\end{tabular}
\end{table}

At low effective temperatures, we do not observe significant temperature corrections. Above $\sim 22\,000$ K, especially for large surface gravities, the temperature corrections can reach up to 3000~K. For $\log{g} = 7.5$ and 8.0, however, the $T_{\rm{eff}}$ corrections become negligible at the highest effective temperatures, where the spectral line forming regions become radiative and therefore equivalent to their 1D counterparts. We do not observe any significant temperature corrections for the V777 Her instability strip \citep{fontaine08} at $\log g = 8.0$. We remind the reader, however, that asteroseismic predictions could be impacted by the significantly different sizes for the 3D convection zones as discussed in Section~\ref{sec23}. It is reassuring that the current ML2/$\alpha$ = 1.25 parameterisation for the optical spectra of DB white dwarfs, which mostly impacts the $T_{\rm{eff}}$ scale, is in reasonable agreement with the 3D simulations.
 
The 1D models tend to significantly overpredict the $\log{g}$ values in the range 14\,000 K $\lesssim T_{\rm{eff}} \lesssim 21\,000$ K for $\log{g} = 7.5$ and 8.0, but this range does extend further to 22\,000 K for $\log{g} = 8.5$ and to 24\,000 K for $\log{g} =9.0$. Above these effective temperatures, the 3D $\log{g}$ corrections are within the BW11 errors. BW11 and \cite{koester_kepler_2015} have shown that DB and DBA white dwarfs in the range 12\,000~K $\lesssim T_{\rm{eff}} \lesssim$ 16\,000~K have larger than expected $\log{g}$ values, with maximum discrepancy between the spectroscopically derived $\log{g}$ values and those predicted by stellar evolutionary models occurring at around 13\,000-14\,000~K. Therefore, our proposed 3D $\log{g}$ corrections are an incomplete solution to this problem. Many studies have attributed the high-$\log{g}$ problem in DB white dwarfs to issues with the line broadening by neutral helium and not with the treatment of convection. Our results provide support for this scenario. Furthermore, a smooth mass versus cooling age distribution for DB stars is expected from evolutionary models. When applying our 3D corrections to the 1D atmospheric parameters determined in BW11 and \cite{koester_kepler_2015} assuming pure-helium atmospheres as a very preliminary assessment, the 3D parameters are not in obviously better or worse agreement with evolutionary models. To fully understand the mass distribution of DB white dwarfs, we believe that 3D simulations with mixed helium and hydrogen compositions must first be calculated, even though \cite{beauchamp_1999_v777_dba} suggest that hydrogen does not significantly impact the atmospheric parameters in the range 14\,000 K $\lesssim T_{\rm{eff}} \lesssim 20\,000$~K. 

A study of the temperature and density stratifications (Figs.~\ref{fig:temp_strat_lhd_atmo_threeD} and \ref{fig:dens_strat_lhd_atmo_threeD}) in the line forming regions (Fig.~\ref{fig:tau_ross_comp}) can be useful to understand the strong predicted 3D corrections at $T_{\rm eff} \sim 18\,000$ K. Fig.~\ref{fig:tau_ross_comp} (top right panel) illustrates that 3D effects on the mean structure are strong enough at this effective temperature that the 3D lines are formed in a significantly narrower range of the atmosphere. Fig.~\ref{fig:dens_strat_lhd_atmo_threeD} shows that in the line forming region, the density is significantly larger in the 3D simulation. Since density correlates with surface gravity, it suggests that a higher gravity 1D structure is necessary to mimic the 3D density stratification, resulting in a negative $\log g$ correction. We note that the spectral lines are formed largely within the convective zone and the 3D effects are especially strong in this regime. We have, therefore, no reason to doubt the accuracy of the 3D simulations or suspect that any approximation we have made would cause spurious 3D effects, especially in light of our success with pure-hydrogen 3D model atmospheres.

For DA white dwarfs, the 3D line cores of the deep lower Balmer lines were shown to be too deep when compared to observed white dwarf spectra \citep{tremblay_2013_spectra}. This 3D prediction is largely caused by adiabatic overshoot at large Peclet number \citep[see, e.g.,][]{newref2,newref1} cooling the 3D structures in the upper layers of the atmosphere, an effect that does not occur in 1D. This discrepancy led us to remove line cores for calculating more robust 3D corrections for DA white dwarfs. For 3D simulations of DB stars we do not observe any obvious issue with the line cores. One reason is that He~I lines are weaker and the cores of the lines do not significantly extend into the overshoot regions. We have tried to remove the line cores from the fits, but this does not meaningfully change the 3D corrections, and thus we suggest keeping the full line shapes in the fitting procedure.

\begin{figure}
	\includegraphics[width=\columnwidth]{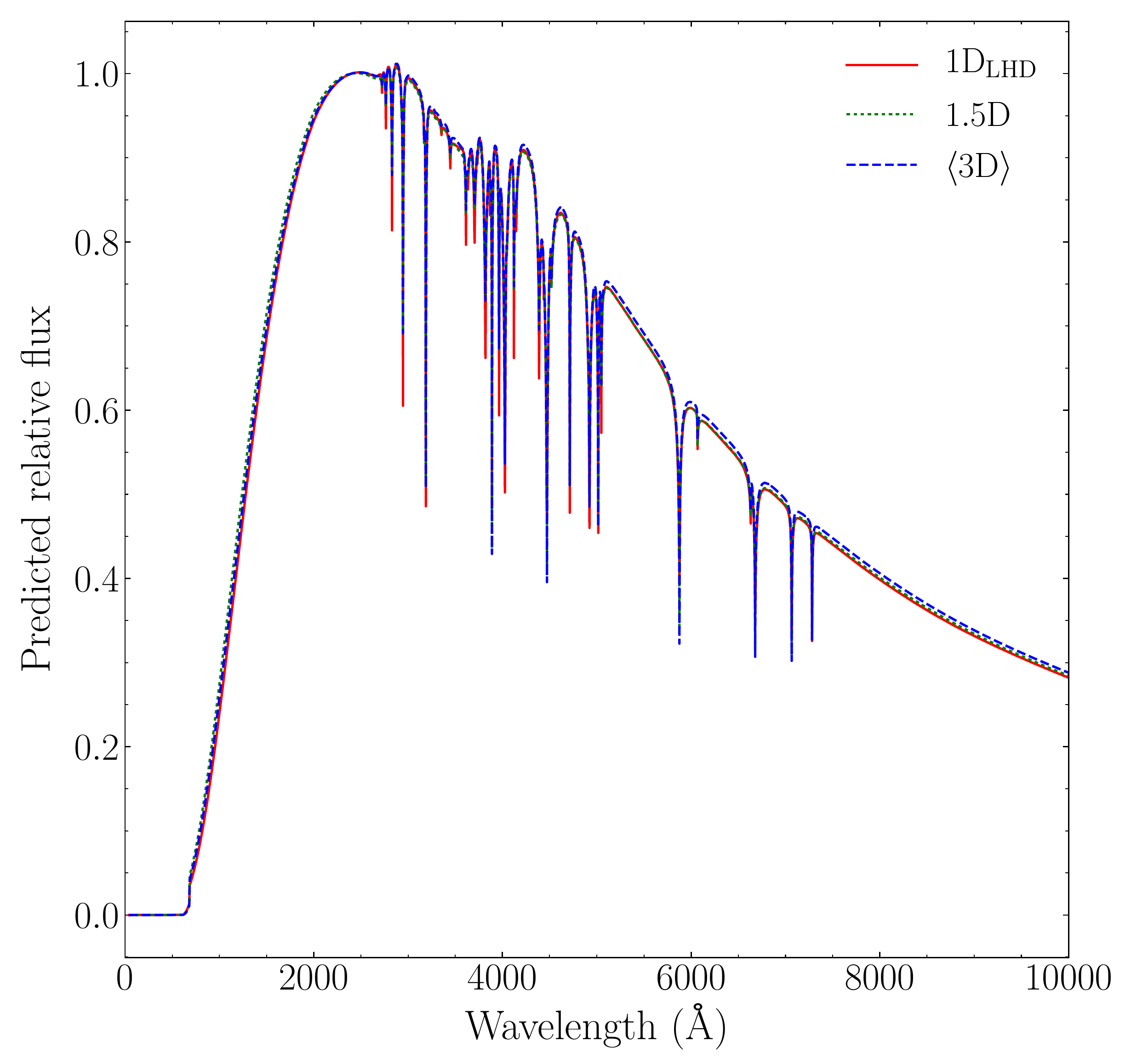}
    \caption{Comparison of 1D LHD (solid red), 1.5D (dotted green) and $\langle$3D$\rangle$ (dashed blue) spectra at $T_{\rm{eff}} = 18\,081$~K and $\log{g} = 8.0$. The spectra have been normalised at 2400~\AA.}
    \label{fig:entire_spectrum_lhd_onefiveD_threeD}
\end{figure}

\begin{figure}
	\includegraphics[width=\columnwidth]{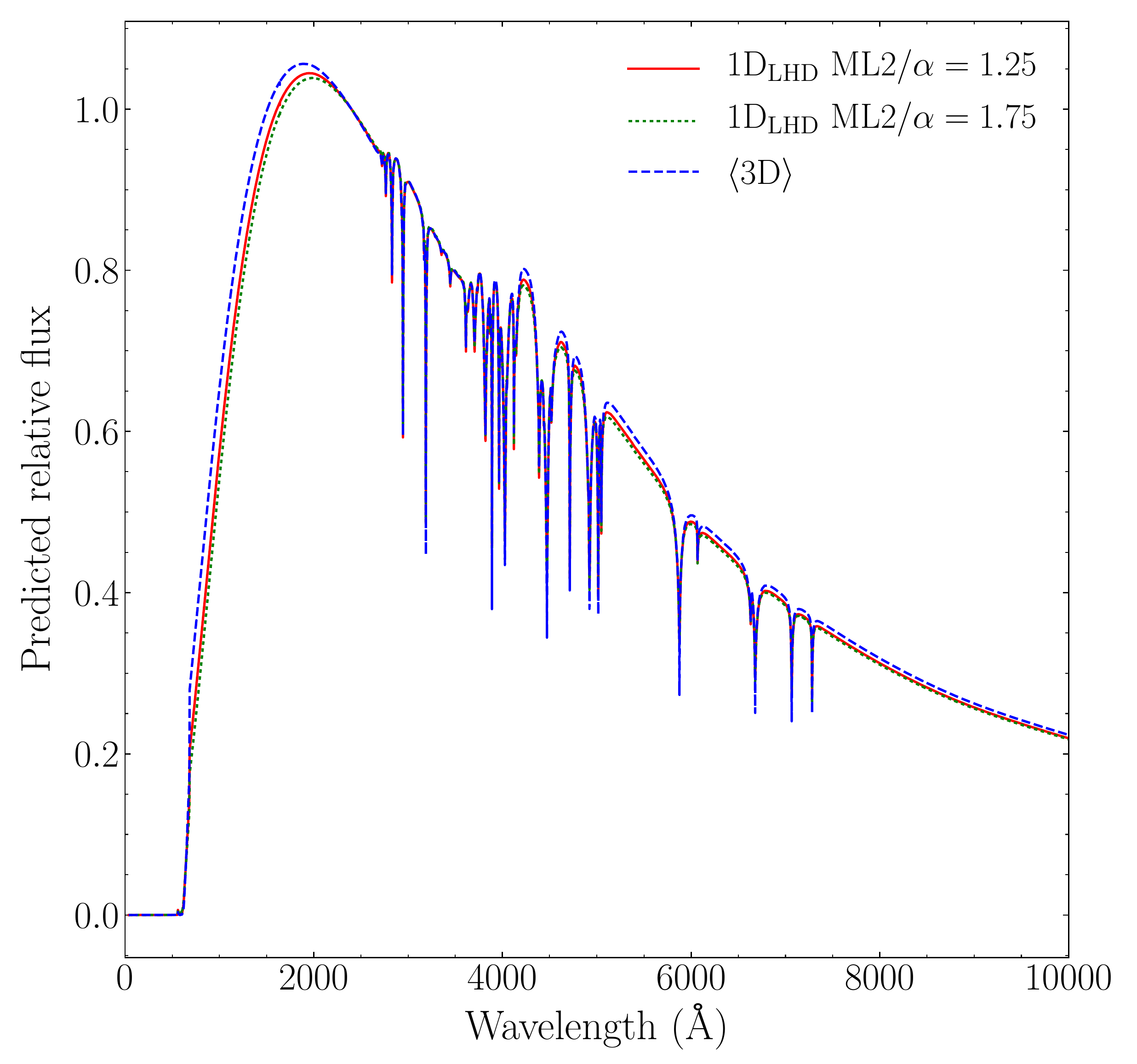}
    \caption{Comparison of a $\langle$3D$\rangle$ (dashed blue) spectrum and 1D LHD spectra computed with ML2/$\alpha$ = 1.25 (solid red) and 1.75 (dotted green) for $T_{\rm{eff}} = 21\,989$~K and $\log{g} = 8.0$. The spectra have been normalised at 2400~\AA.}
    \label{fig:alpha_175_125_spec_comp}
\end{figure}

Fig.~\ref{fig:entire_spectrum_lhd_onefiveD_threeD} compares the normalised 1D LHD, 1.5D and $\langle$3D$\rangle$ spectra for $\log{g} = 8.0$ and $T_{\rm{eff}} = 18\,081$~K, where the largest 3D corrections for $\log{g}$ are observed (for $\log{g} = 8.0$ models). We find that all predicted spectra are very similar in terms of the broadband fluxes from the near-UV to the near-infrared. This suggests once again that the $\langle$3D$\rangle$ approximation is adequate, but also that 3D corrections are unnecessary for calculating broadband photometric fluxes in this regime. In Fig.~\ref{fig:alpha_175_125_spec_comp} we compare our $\langle$3D$\rangle$ spectrum at $\log{g} = 8.0$, $T_{\rm{eff}} = 21\,989$~K with 1D LHD spectra using both ML2/$\alpha$ = 1.25 and 1.75. In this regime, BW11 have found (see their Figure~15) that a mixing-length of ML2/$\alpha$ = 1.75 provides a better agreement between the optical and near-UV temperatures, while a value of ML2/$\alpha$ = 1.25 results in a smoother mass distribution as a function of $T_{\rm eff}$. They attribute this behaviour to a potential shortcoming of the mixing-length theory. It is difficult to conclude yet about the possible improvements of a 3D spectral analysis, since 3D $T_{\rm{eff}}$ corrections are fairly mild in this regime and the predicted near-UV fluxes are all very similar in a relative sense.

\subsection{Sensitivity to input parameters} \label{sec:numerics}

$\langle$3D$\rangle$ thermal structures of DA white dwarfs show little sensitivity to the input numerical parameters, which include the grid resolution, artificial viscosity, geometrical dimensions, and numerical schemes for the hydrodynamics solver (see Table 3 of \cite{tremblay_2013_3dmodels} for more detail). This work relies on the same numerical setup and a change of the gas composition is not expected to have a significant impact on the precision of the thermal structures. We conclude that this earlier investigation puts forward a numerical test for our DB models as well. This does not rule out, however, that there are untested numerical setups \citep[e.g., very large grid sizes, see][]{kupka18} that could still have an effect on our results. These experiments were performed for a characteristic closed bottom simulation ($T_{\rm eff} = 12,000$~K, $\log g = 8.0$, pure-hydrogen). In this section we expand on these numerical experiments by quantifying how the open bottom boundary condition impacts our derived corrections.

\cite{Grimm_Strele_2pressurescale} have shown that vertical boundary conditions can influence layers located two pressure scale heights above or below them. We extended two models at $\log{g} = 8.0$, and initially with $T_{\rm{eff}} = 12\,020$ and 18\,081 K, by adding 60 and 50 more grid points to the bottom of the two simulations, respectively. The former case is the shallowest model at $\log g = 8.0$ with a total of 5 pressure scale heights. For both simulations we only focus on the lower boundary as the top of each simulation is more than 3 pressure scale heights above the top of the spectral line forming region. These new simulations are run for 10 more seconds, and we make sure they have been properly relaxed using the tests described in Section~\ref{sec:numsetup}. 12 snapshots over the last quarter of the simulations are used to calculate the mean structures and synthetic spectra. The two new synthetic spectra are fitted with the 1D LHD grid to derive 3D corrections.

\begin{figure}
	\includegraphics[width=\columnwidth]{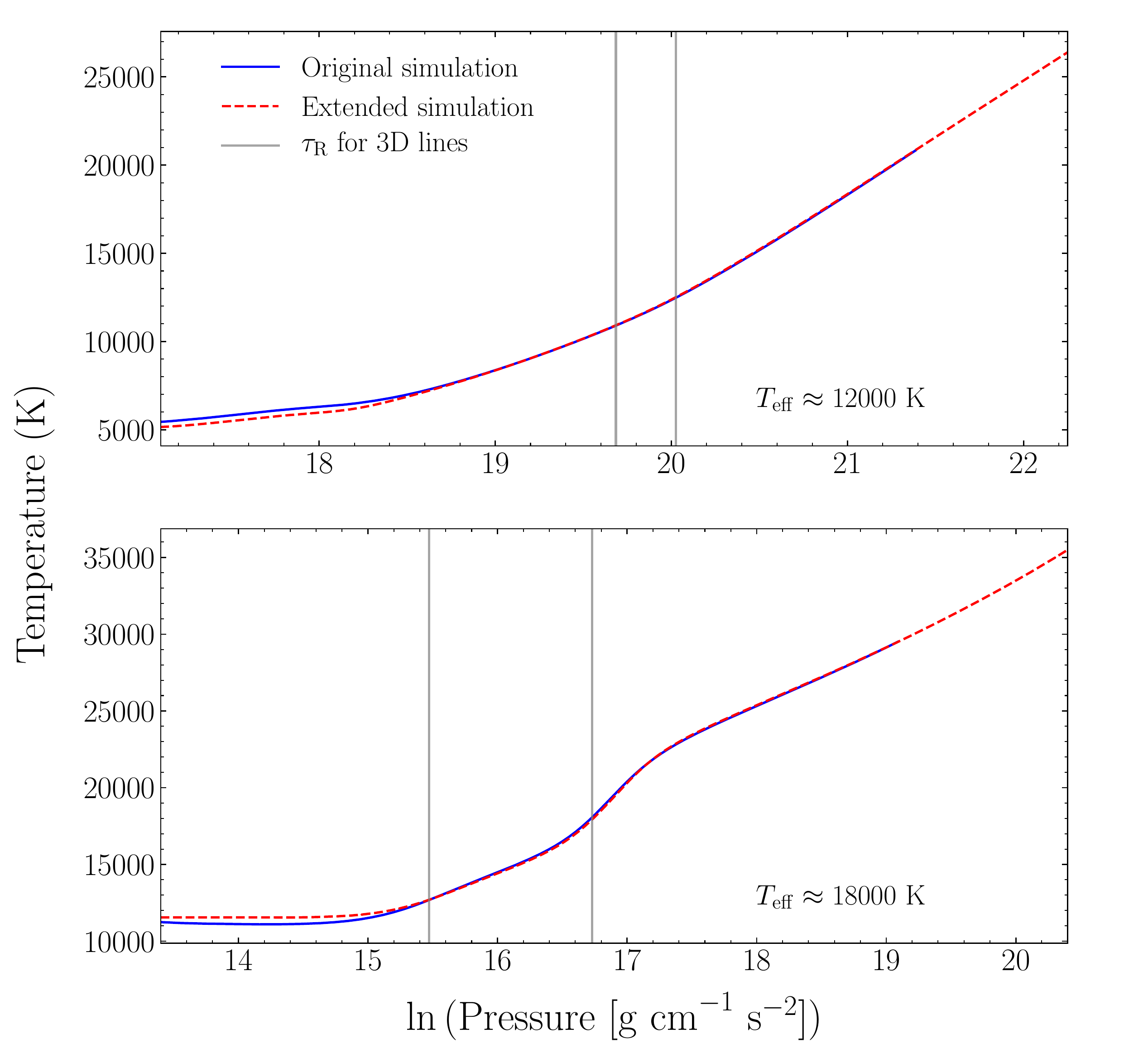}
    \caption{The temperature and pressure stratifications of the original (solid blue) and extended (dashed red) simulations for $\log{g} = 8.0$ models with $T_{\rm{eff}} \approx 12\,000$ and 18\,100 K. The spectral line forming regions are indicated by solid grey lines. Note that the 3D simulations extend deeper into the upper layers than shown here.}
    \label{fig:extended_hp_comp}
\end{figure}

In Fig.~\ref{fig:extended_hp_comp} we compare the temperature and pressure stratifications between the original and extended simulations. We find that the $\langle$3D$\rangle$ structure at $\approx$ 12\,000~K does not change significantly with the extended simulation. The 3D spectroscopic corrections are well within fitting errors. Convection is very adiabatic everywhere in the simulation and we hypothesize that the mean stratification is rather insensitive to the treatment of convection (either in 1D or 3D). The standard and extended $\approx$ 18\,000~K simulations differ marginally in the line forming regions according to Fig.~\ref{fig:extended_hp_comp}. The shift in the 3D $\log g$ correction is similar to the typical external observational errors ($\approx 0.05$ dex). The original simulation was already deep in terms of the number of pressure scale heights between the photosphere and the bottom boundary, and therefore the difference may not be directly caused by the change in the bottom boundary condition.

\subsection{Application to observations}

\begin{figure*}
    \centering
	\begin{subfigure}[b]{2\columnwidth}
		\centering
        \includegraphics[trim={2cm 2cm 3cm 5cm}, clip, angle=270, width=0.8\columnwidth]{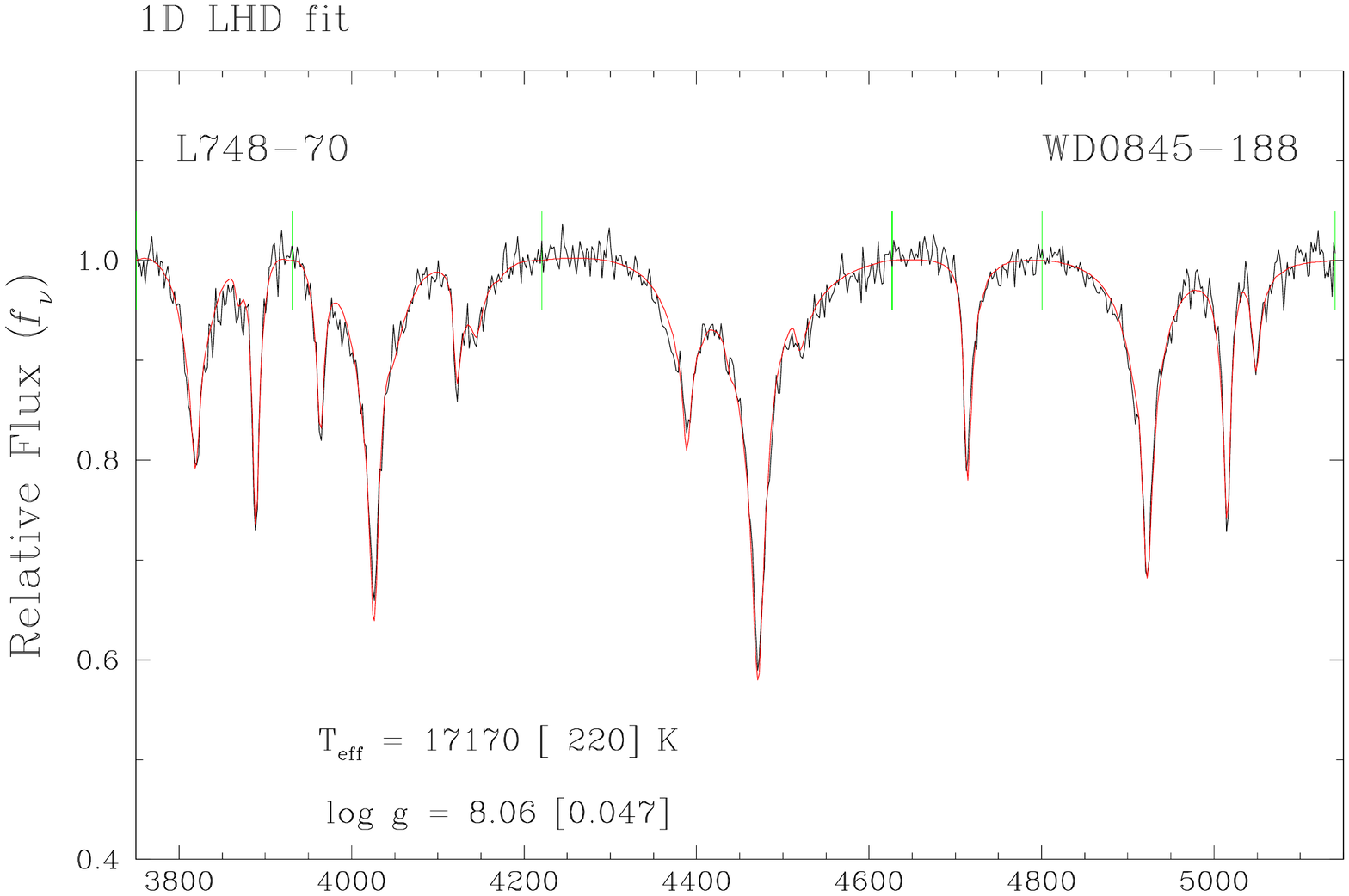}
        \label{fig:fitfinal_0845-188_1d}
    \end{subfigure}
    \begin{subfigure}[b]{2\columnwidth}
    	\centering
        \includegraphics[trim={2cm 2cm 2cm 5cm}, clip, angle=270, width=0.8\columnwidth]{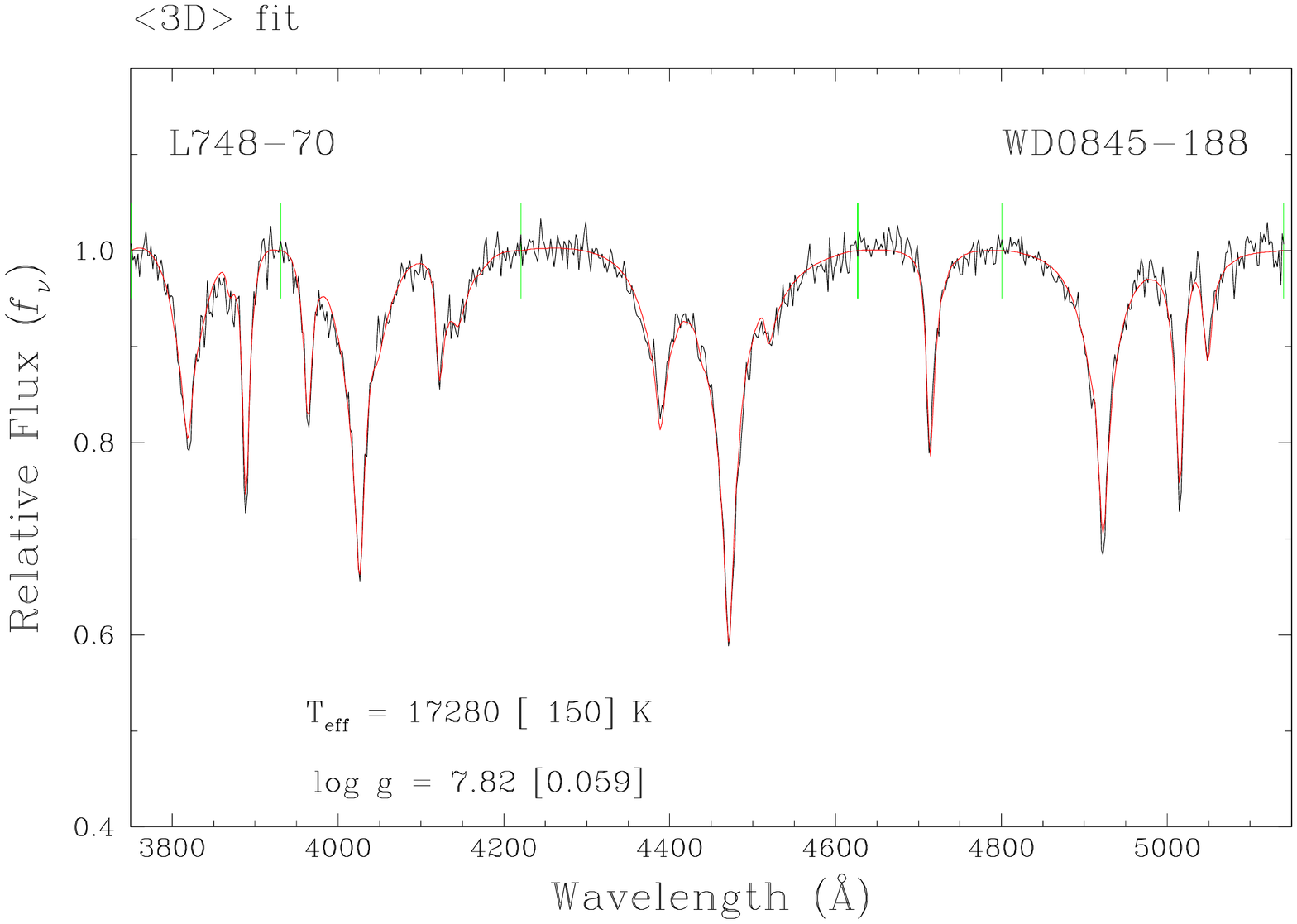}
        \label{fig:fitfinal_0845-188_3d}
    \end{subfigure}
    \caption{Example 1D LHD (top panel) and $\langle$3D$\rangle$ (bottom panel) fits to the spectrum of the DB white dwarf WD~0845$-$188. The continuum flux
is fixed to unity by a fitting function at predefined wavelength points shown as green tick marks in the panels (see BW11). The best fit atmospheric parameters assuming a pure-helium composition are identified on the panels.}\label{fig:wd_fit_plots}
\end{figure*}

Fig.~\ref{fig:wd_fit_plots} shows 1D LHD and $\langle$3D$\rangle$ fits to WD0845$-$188, a selected DB white dwarf from BW11 with hydrogen abundance small enough to assume pure-helium composition \citep{bergeron_WD0845_2015}. Fitting with $\langle$3D$\rangle$ spectra lowers the $\log{g}$ by 0.24 dex, in line with the corrections proposed in Table~\ref{tab:3d_corr}. However, the $T_{\rm{eff}}$ difference does not exactly match the corrections proposed in Table~\ref{tab:3d_corr}, but since the correction is of the same order as the internal errors we believe this inconsistency to be negligible. 

If we fit WD0845$-$188 with 1D ATMO instead of 1D LHD, we recover parameters that are almost in complete agreement to LHD fitted parameters, reinforcing what is shown in Fig.~\ref{fig:lhd_atmo_corr_logg80}, i.e. the difference between the 1D structures calculated from these two 1D codes are negligible at the given $\log{g}$ and $T_{\rm{eff}}$.

For this particular DB white dwarf, the $\chi^2$ is marginally smaller in the $\langle$3D$\rangle$ case compared to the 1D LHD fit. However, looking at the whole BW11 sample excluding DBA white dwarfs, we do not find an obvious preference for either model grid, suggesting that fits are of equivalent quality on average. This is in line with our earlier finding that there is no obvious line core problem for DB white dwarfs in comparison to DA stars. This suggests that the next step is to calculate a grid of mixed He/H 3D atmospheres and revisit earlier spectroscopic analyses.

\section{Conclusions} \label{sec:conc}

We have presented the first-ever 3D radiation-hydrodynamics simulations of DB white dwarf atmospheres and discussed them in terms of the 3D effects on synthetic spectra. 
Briefly examined were the significant differences between these new 3D models and their previously available 1D counterparts in terms of the temperature and density stratifications. This distinction arises from the different models of convection; the 3D treatment derived from first principles and the more approximate mixing-length theory in 1D. Our 3D simulations are not without approximations either, but these issues can be largely overcome when computing 3D corrections with carefully selected reference 1D models. In our case, the sister-code of CO$^5$BOLD, LHD, was used, which treats opacity binning and scattering in the exact same fashion as CO$^5$BOLD.

The 3D corrections on the atmospheric parameters were constrained by using both 1.5D and $\langle$3D$\rangle$ spectra, which represent the two extremes of enhancing or neglecting the 3D fluctuations, respectively. Corrections found with either method are similar, and the differences are within typical fitting uncertainties, suggesting that full 3D spectral synthesis is not required. The $\langle$3D$\rangle$ spectra, drawn from $\langle$3D$\rangle$ structures averaged over constant optical depth, have thus been used to estimate 3D corrections for pure-helium atmosphere white dwarfs. We find that current 1D synthetic spectra, under the ML2/$\alpha$ = 1.25 parameterisation of the mixing-length theory, overpredict $\log{g}$ in the range 12\,000~K $\le T_{\rm{eff}} \le 23\,000$~K by as much as 0.4 dex. It is a surprising result since DB white dwarf parameters have not been reported to be erroneous in this $T_{\rm{eff}}$ range.

Photometric fits using {\it Gaia} Data Release 2 are expected to provide independent masses for all known DB stars, giving us a better description of the shortcomings in the line broadening or current 1D and 3D model atmospheres. The next step will be to fully revisit the earlier 1D spectroscopic analyses by computing a grid of mixed He/H 3D simulations. This will account for the hypothesis that most if not all helium-rich atmosphere white dwarfs have hydrogen traces \citep{koester_kepler_2015}. The 1D envelopes will also be re-calibrated with updated mixing-length parameters that are able to reproduce the size of the 3D convection zones as in \cite{tremblay15}.
 
\section*{Acknowledgements}

This project has received funding from the European Research Council (ERC) under the European Union's Horizon 2020 research and innovation programme (grant agreement No 677706 - WD3D). H.G.L. acknowledges financial support by the Sonderforschungsbereich SFB\,881 ``The Milky Way System'' (subprojects A4) of the German Research Foundation (DFG).




\bibliographystyle{mnras}
\bibliography{aamnem99,3d_db_elena_cukanovaite_et_al} 








\bsp	
\label{lastpage}
\end{document}